\documentclass[11pt]{article}
\usepackage{latexsym,amssymb,amsmath,graphicx,epsfig,amsthm}
\newtheorem{teo}{Theorem}
\newtheorem{deft}{{\bfseries{Definition}}}
\newtheorem{lem}{Lemma}
\newtheorem{example}{Example}
\usepackage{bm}
\usepackage{natbib}
\usepackage{caption}
\usepackage{subcaption}
\usepackage{color}
\usepackage{xcolor}
\usepackage{verbatim}
\usepackage{booktabs}
\usepackage{multirow}
\usepackage[colorlinks=true,linkcolor=red, citecolor=blue]{hyperref}
\usepackage{doi}

\usepackage{dsfont}
\title{\bf Comparing two spatial variables with the probability of agreement}
\author{
Jonathan Acosta$^1$, Ronny Vallejos$^2$, Aaron M. Ellison$^{3,4}$, Felipe Osorio$^2$, M\'{a}rio de Castro$^5$ \\ \\
$^1$Departamento de Estadística,
Pontificia Universidad Católica de Chile\\
$^2$Departamento de Matemática, Universidad Técnica Federico Santa María, Chile\\
$^3$Harvard University Herbaria, Harvard University, Cambridge, MA, USA\\
$^4$Sound Solutions for Sustainable Science, Boston, MA, USA\\
$^5$Instituto of Ci\^encias Matemáticas e de Computa\c c\~ao\\Universidade de S\~ao Paulo, S\~ao Carlos, Brazil
}

\date{}
\begin{document}
\maketitle

ORCIDs:
\begin{itemize}
    \item[] Jonathan Acosta: \href{https://orcid.org/0000-0001-6323-9746}{0000-0001-6323-9746}
    \item[] Ronny Vallejos: \href{https://orcid.org/0000-0001-5519-0946}{0000-0001-5519-0946}
    \item[] Aaron M. Ellison: \href{https://orcid.org/0000-0003-4151-6081}{0000-0003-4151-6081}
    \item[] Felipe Osorio: \href{https://orcid.org/0000-0002-4675-5201}{0000-0002-4675-5201}
    \item[] Mario de Castro: \href{http://orcid.org/0000-0001-8685-9470}{0000-0001-8685-9470}
\end{itemize}

\begin{abstract}
Computing the agreement between two continuous sequences is of great interest in statistics  when comparing two instruments or one instrument with a gold standard. The probability of agreement (PA) quantifies the similarity between two variables of interest, and it is useful for accounting what constitutes a practically important difference. In this article we introduce a generalization of the PA for the treatment of spatial variables. Our proposal makes the PA dependent on the spatial lag. As a consequence, for isotropic stationary and nonstationary spatial processes, the conditions for which the PA decays as a function of the distance lag are established. Estimation is addressed through a first-order approximation that guarantees the asymptotic normality of the sample version of the PA. The sensitivity of the PA is studied for finite sample size, with respect to the covariance parameters. The new method is described and illustrated with real data involving autumnal changes in the green chromatic coordinate ($G_{cc}$), an index of ``greenness'' that captures the phenological stage of tree leaves, is associated with carbon flux from ecosystems, and is estimated from repeated images of forest canopies.

\vspace{1mm}

\end{abstract}

\section{Introduction}
The comparison of two sequences is a fundamental problem in several scientific disciplines, and it can be addressed in many different ways. For example, the Student's \textit{t}-test, the correlation coefficient, and the Wilcoxon two-sample rank test are three techniques that, under different assumptions, provide information about some aspects of comparisons between two independent populations \citep{Lin:2012}. When the goal is to measure the agreement between two variables to validate an assay, a process, or a newly developed instrument, it may be relevant to evaluate whether its performance is concordant with other existing ones or a ``gold standard’’ \citep{Lin:2002}.

The probability of agreement (``PA'') measures the level of agreement between two continuous sequences. It was first introduced in a series of papers by Nathaniel Stevens et al. \citep{Stevens:2017a, Stevens:2017b, Stevens:2018, Stevens2:2020, Stevens:2020} as an alternative approach to some existing agreement measures, including the concordance correlation coefficient between measurements generated by two different methods \citep{Lin:1989} and its extensions. Subsequently, \cite{Leal:2019} studied the PA in a context of local influence. \cite{De Castro:2021} developed Bayesian PA methods to compare measurement systems with either homoscedastic or heteroscedastic measurement errors. We note that the literature on PA is extensive and many relevant uses for it in different scenarios have been published, but there is as yet no single definition for PA. A salient example is by \cite{Ponnet:2021}, who adapted the probability of concordance or C-index to the specific needs of a discrete frequency and severity model that typically is used during the technical pricing of a non-life insurance product.

In this paper, we generalize the PA \citep[sensu][]{Lin:2002, Stevens:2017a, Stevens:2017b} to the case of two georeferenced sequences in the plane. This generalization makes the PA dependent on a spatial lag, similar to how the variogram and covariance functions are used in spatial statistics. Specific conditions for the variance of the difference between the two sequences are imposed so that the PA is a decreasing function when the norm of the spatial lag increases. This monotonic feature of the PA can be established easily for the bivariate Mat\'{e}rn and Wendland covariance functions. We then extend the PA for the case of spatiotemporal processes so that two images of the same scene taken at different times can be compared. This extension is motivated by the need to track temporal changes in spatial patterns. In our spatiotemporal extension of the PA, we consider that the process includes a temporal and, perhaps, a spatial trend, and random noise. The resulting process is nonstationary in the mean and is flexible enough to account for a number of different trends in time.

Estimation of the PA is addressed via plug-ins and the delta method, assuming that the estimates of the parameters of the covariance function exist and are asymptotically normal \citep{Mardia:1984, Acosta:2018}. A simple expression for the asymptotic variance of the sample PA is also derived. To assess the properties of the PA for finite sample sizes, we carried out two numerical experiments: a sensitivity study that illustrates that PA decreases as a function of the norm of the spatial lag, and a Monte Carlo simulation study to estimate the PA and the relevant parameters of two spatial covariance models. Finally, we apply our spatial PA to a temporal sequence of images of a forest canopy and estimate the PA of two images taken of the same scene at different times. Images like these are used routinely to identify seasonal changes in the unfolding, maturation, and senescence (with accompanying fall colors) of individual trees and entire forest canopies, and to estimate fluxes of carbon, water vapor, and other gases between the forest and the atmosphere \citep[e.g.,][]{Richardson:2018}.

In Section \ref{sec:background} we give some additional, albeit brief, background on PA. In Sections \ref{sec:concordance} and \ref{sec:saptiotemporal} we introduce the idea of PA for spatial processes and establish the main theoretical results for stationary processes (Section \ref{sec:concordance}) and spatiotemporal processes (Section \ref{sec:saptiotemporal}). Section \ref{sec:estimation} discusses estimation of PA, which is then illustrated with numerical simulation experiments (Section \ref{sec:numerical_exp}) and an empirical example (Section \ref{sec:application}). We conclude with an outline for future research and developments in this area (Section \ref{sec:conclusions}). Proofs of the four theorems and two lemmas used in the paper are given in the Appendix.

\section{Background and preliminaries}\label{sec:background}
Following \cite{Leal:2019}, we assume that $\{(X_{11}, X_{21}),\ldots, (X_{1n}, X_{2n})\}$ is a random sample from a bivariate normal distribution with mean vector $\bm \mu$ and covariance matrix $\bm \Sigma $. Then, a method to quantify the degree of agreement between the variables $X_1$ and $X_2$ relies on the differences between their corresponding values: $$D_i = X_{1i}-X_{2i}, \hspace{3mm} i=1,\ldots,n.$$ The probability of agreement is defined as
 \begin{equation}\label{eq:concordance}
\psi_c=\text{P}(\ |D_i|\leq c), \ \ c>0,
\end{equation}
where $c$ denotes the maximum acceptable difference from a practical perspective. In such a case the interval $(-c,c)$ is often called the ``clinically acceptable difference'' (CAD).

Because of the normality assumption, the PA as given in Equation \eqref{eq:concordance} takes the form
\begin{equation}\label{eq:concor2}
\psi_c=\Phi\left(\frac{c-\mu_D}{\sigma_D}\right)- \Phi\left(-\frac{c+\mu_D}{\sigma_D}\right),
\end{equation}
where $\Phi(\cdot)$ denotes the cumulative distribution function of the standard normal, $\mu_D = \mu_1 - \mu_2$,  and $\sigma_D^2= \sigma_{11} + \sigma_{22} - 2\sigma_{12}$.
How large the PA should be to consider the variables interchangeable is up to the practitioner; \cite{Stevens:2017b} suggest using $\psi_c\geq 0.95$ as a guideline.

Under the assumption of normality, inference for $\bm \mu$ and $\bm \Sigma$ can be addressed via maximum likelihood (ML) \cite[][\S3.2]{Anderson:2003}. By substituting such estimates into Equation \eqref{eq:concor2}, an ML estimate for $\psi_c$, denoted by $\widehat{\psi}_c$, can be obtained. Under mild assumptions, \cite{Leal:2019} established the asymptotic normality of $\widehat{\psi}_c$, which relies on the asymptotic distribution of the ML under normality and the delta method. It is then straightforward to estimate approximate confidence intervals and test hypotheses about $\psi_c$.

\section{Probability of agreement for stationary processes}\label{sec:concordance}
In this section we introduce the PA in the context of georeferenced variables.

Let $\bm Z(\bm s)=(X(\bm s), Y(\bm s))^{\top}$ be a bivariate second-order stationary random field with $\bm s, \bm h \in \mathds{R}^2$, mean $(\mu_X, \mu_Y)^{\top}$, and covariance function
$$\bm C(\bm h)=\left(\begin{matrix} C_X(\bm h) & C_{XY}(\bm h)\\
C_{YX}(\bm h) & C_{Y}(\bm h)
\end{matrix} \right),$$
where
\begin{align*}
 C_{X}(\bm h) &= \text{cov}[X(\bm s),X(\bm s+\bm h)],  \\
 C_{Y}(\bm h) &= \text{cov}[Y(\bm s),Y(\bm s+\bm h)],  \\
 C_{XY}(\bm h)&= C_{YX}(\bm h)= \text{cov}[X(\bm s),Y(\bm s+\bm h)],
\end{align*}
and $(\cdot)^{\top}$ means  transposition.
Define the difference
\begin{equation}\label{eq:spatdiff}
D(\bm s,\bm h)=X(\bm s) -Y(\bm s+\bm h).
\end{equation}
This difference measures the discrepancy between the processes when there is a separation vector equal to $\bm h$ between them.
Assume that $\bm Z(\bm s)=(X(\bm s),Y(\bm s))^\top, ~ \bm s\in D\subset \mathbb{R}^2$ is a Gaussian process with mean  $\bm \mu= (\mu_X,\mu_Y)^\top$  and covariance function $\bm C(\bm h)$, $\bm h \in D$. Then,
$$D(\bm s,\bm h)\sim \mathcal{N}(\mu_D, \sigma^2_D(\bm h)),$$
where $\mu_D=\mu_X-\mu_Y$ and $\sigma^2_D=C_X(\bm 0)+C_Y(\bm 0 )-2C_{XY}(\bm h).$
Then, the PA between processes $X(\bm s)$ and $Y(\bm s+\bm h)$ is
\begin{equation}\label{eq:spatconcordance}
\psi_c(\bm h)=\text{P}[\ |D(\bm s, \bm h)|\leq c], \ \ c>0,
\end{equation}
which assumes the form
\begin{equation}\label{eq:psi_c}
\psi_c(\bm h)= \Phi\left(\frac{c-\mu_D}{\sigma_D(\bm h)}\right)- \Phi\left(-\frac{c+\mu_D}{\sigma_D(\bm h)}\right),
\end{equation}
where $\Phi(\cdot)$ is as in Equation \eqref{eq:concor2}. If we also assume isotropic processes, the probability (Equation \eqref{eq:spatconcordance}) can be plotted as a function of $\|\bm h\|$ (where $\| \cdot \|$ is the Euclidean norm in $\mathbb{R}^2$) in a similar way as the covariance function is plotted for several parametric processes. In that case, the PA (Equation \eqref{eq:concordance}) is obtained as a particular case of Equation \eqref{eq:spatconcordance}. In fact, $\psi_c=\psi_c(\bm{0})$.

We first consider  the Mat\'ern covariance function \citep{Matern:1986} to illustrate Equation \eqref{eq:psi_c}. This function is widely used in spatial statistics because of its theoretical properties and its flexibility for modeling local behavior of spatial correlations \citep{Stein:1999, Guttorp:2006}. It also is used in machine learning and with neural networks \citep{Rasmussen:2006}. The Mat\'ern covariance function takes the form
\begin{equation}\label{eq:mat}
M(\bm h, \nu, a)=\frac{2^{1-\nu}}{\Gamma(\nu)}(a \| \bm h \|)^{\nu}K_{\nu}(a \|\bm h \|),
\end{equation}
where $K_{\nu}(\cdot)$ is a modified Bessel function of the second kind,   $a>0$ is a parameter that controls the rate of decay of the correlation, and $\nu>0$ is the smoothing parameter that is related to the behavior of the correlation near the origin. A special case of the Mat\'ern  function is when  $\nu= m + 1/2, \  m \in \mathbb{N}_0$. Then,
\begin{align}\label{eq:mat_int}
 M(\bm h,m+1/2,a)= \exp(-a \|\bm h\|) \sum_{k=0}^{m} \dfrac{(m+k)!}{(2m)!} \begin{pmatrix} m\\k \end{pmatrix} (2a \|\bm h\|)^{m-k}
\end{align}
By choosing  $m=0$, we have the simplest form  $ M(\bm h,1/2,a) = \exp(-a \|\bm h\|)$.
In the sequel, when dealing with isotropic models, we will use $h$ for $\|\bm h\| \in \mathbb{R}^2.$

For a bivariate Gaussian random field  the Mat\'ern covariance function has been extended \citep{Gneiting:2010} and defined as
 \begin{align}
C_X(h)&=\sigma_X^2 M(h,\nu_X,a_X), \label{eq:mat1}\\
C_Y(h)&=\sigma_Y^2 M(h,\nu_Y,a_Y),\label{eq:mat2}\\
C_{XY}(h)&=\rho_{XY}\sigma_X \sigma_YM(h,\nu_{XY},a_{XY}),\label{eq:mat3}
\end{align}
where $\sigma_X^2>0, \sigma_Y^2>0,$ and $\rho_{XY}$ is the co-located correlation coefficient between $X(\cdot)$ and $Y(\cdot)$. Specific conditions for the parameters are required so that the covariance model described in equations \eqref{eq:mat1}-\eqref{eq:mat3} is positive definite. In this model we have that $\sigma_D^2(h)=\sigma_X^2+\sigma_Y^2-2\rho_{XY}\sigma_X \sigma_YM(h,\nu_{XY},a_{XY})$.

For illustrative purposes, consider  $\sigma_X=1$, $\sigma_Y=2$, $\sigma_{XY}=1.8$, $a_{XY}=2$, $\rho_{XY}=0.9$ and $\nu = \nu_{XY} = \lbrace 0.5, 1.5, 2.5 \rbrace$. We plot $\psi_c(h)$ versus $h$ for
$ h \in \{0,1, \ldots,15\}$,   $c=\lbrace  1.5, 2, 2.5  \rbrace$,  and using the Mat\'ern covariance function (Figure 1 in Supplementary Material). In all cases, $\psi_c(h)$ decreases as a function of $h$ and the curves decay more rapidly to zero as $\nu$ decreases. This is a consequence of the monotonic property of the Mat\'ern covariance as shown in the following example.


\begin{example}\label{ex:matern1}
\normalfont
Let $\bm Z(\bm s)$ be a bivariate second-order stationary Gaussian process with the Mat\'ern covariance function given in equations \eqref{eq:mat1}-\eqref{eq:mat3}. Assume that $\nu_{XY}=m+\frac{1}{2}$ and,
without loss of generality, assume that the Gaussian process has mean zero and $a=1$ in Equation \eqref{eq:mat_int}. Then,
\begin{align*}
\psi_c(h)&=\Phi\left(\frac{c}{\sigma_D(h)}\right)- \Phi\left(-\frac{c}{\sigma_D(h)}\right)
=2\Phi\left(\frac{c}{\sigma_D(h)}\right)-1.
\end{align*}
This implies that
\begin{align}\label{eq:aux1}
\psi^{\prime}_c(h)&=2 \varphi \left(\frac{c}{\sigma_D(h)}\right)\left(-\frac{c}{\sigma_D(h)}\right)\frac{1}{2\sqrt{\sigma_X^2+\sigma_Y^2-\rho_{XY}\sigma_X\sigma_YM(h,\nu_{XY},a_{XY})}}\notag \\
& \quad \times[- \rho_{XY}\sigma_X\sigma_Y M^{\prime}(h,\nu_{XY},a_{XY})],
\end{align}
where $\varphi(\cdot)$ is the probability density function of a standard normal random variable. Because the  right hand side of Equation \eqref{eq:aux1} is positive, to prove that $\psi_c(h)$ is decreasing as a function of $h$,
it is enough to get the sign of the derivative of $M(h,\nu_{XY},a_{XY})$ with respect to $h$.
We write  $M(\cdot, \cdot,\cdot)$ in Equation \eqref{eq:mat_int}  as
$M(h,m+\frac{1}{2},1)=M(h)=\exp(-h)P_m(h),$
where $P_m(x)=\sum_{n=0}^m a_n x^n,$ with $a_n=\frac{2^n}{n!}{m \choose n}/ {\ 2m \choose n}$ \citep{Acosta:2018}. Then
$M^{\prime}(h)=[P^{\prime}_m(h)-P_m(h)]\exp(-h).$ Consequently,
solving the equation $M^{\prime}(h)=0$ is equivalent to solve $P_m^{\prime}(h)=P_m(h).$ Because
$$P_m^{\prime}(h)=a_1+2a_2h+\cdot+ma_mh^{m-1},$$
we have that
$$Q(h):=P_m^{\prime}(h)-P_m(h)=\sum_{n=0}^{m} b_nh^n,$$
where $b_n=(n+1)a_{n+1}-a_n, n=0,\cdots,m-1,$ and $b_m=-a_n.$ Consider
$$a_n=\frac{2^n m! (2m-n)!}{n! (m-n)!}.$$
Note that
\begin{align*}\label{eq:aux2}
b_n&=\frac{-2^n m! (2m-n-1)!)}{n! (m-n-1)! (2m)!}\left(\frac{2m-n}{m-n}-2\right)\\
&=\frac{-2^n m! (2m-n-1)!)n}{n! (m-n-1)! (2m)!(m-n)}<0,
\end{align*}
for $n=0,\cdots,m-1,$ and $m\geq 1.$ Furthermore, $a_m=\frac{2^m m!}{(2m)!}>0.$
Thus $b_0=0$ and $b_n<0,$ for $n=0,\cdots,m-1,$ and $m\geq 1.$ It then follows that
$$M^{\prime}(h)=0  \iff h=0.$$
Moreover,  $M^{\prime}(h)<0$ for $h>0$, because $M^{\prime}(h)=Q(h)e^{-h}$, and $Q$ is a polynomial with  negative coefficients. Therefore, we have explicitly shown that $\psi_c(h)$ is a decreasing  function of $h$.
\end{example}

\bigskip

It should be noted that $\psi_c(\|\bm h\|)$ will not necessarily be a decreasing function of $\|\bm h\|$. As an example, consider a bivariate process with mean $(\mu,\mu)^{\top}$ and a separable covariance function $\bm C(h)$, where $C_X(h)=C_{Y}(h)=\sigma^2 (\phi/h) \sin(h/\phi),$ and $C_{XY}(h)=\rho_{XY}\sigma^2(\phi/h)\sin(h/\phi)$. For simplicity set $\sigma^2=1$ and $\phi=1$. Then, $\sigma_D^2(h)=2\left(1-\rho_{XY} \sin(h)/h\right).$ Clearly, for $c=1$ and $\rho_{XY}=\frac{1}{2}$, $\psi_c(\pi/2)=0.6082,$ $\psi_c(3\pi/2)=0.4986,$ and $\psi_c(5\pi/2)=0.5351,$ hence $\psi_c(h)$ is not decreasing in $h$.

However, for certain parametric models, as in Example \ref{ex:matern1}, $\psi_c(\|\bm h\|)$ is a monotonic function.  A sufficient condition for a parametric covariance model that ensures that $\psi_c(\|\bm h\|)$ is a decreasing monotonic function is given by Theorem \ref{th:main} (the proof of this and subsequent theorems are in the Appendix).

\begin{teo}\label{th:main}
Suppose that $\psi_c(\|\bm h\|)$ is as in \eqref{eq:psi_c}.  If $\sigma_{D}(\|\bm h\|)$ is an increasing function of $\|\bm h\|$, then $\psi_c(\|\bm h\|)$ is a decreasing function of $\|\bm h\|$.
\end{teo}

Theorem \ref{th:monotonia_matern} shows that for the bivariate Mat\'{e}rn covariance function defined in Equations \eqref{eq:mat1}-\eqref{eq:mat3},   $\sigma_D^2(\|\bm{h}\|)$ is an increasing function of $\|\bm{h}\|$.

\begin{teo}\label{th:monotonia_matern}
Suppose that $\sigma_D^2(\|\bm{h}\|)$ is obtained using the Mat\'{e}rn covariance model and assume that $\rho_{XY}\geq0$. Then $\sigma_D(\|\bm{h}\|)$ is an increasing function of $\|\bm{h}\|$. In consequence, the conditions of Theorem \ref{th:main} are satisfied and  $\psi_c(\|\bm h\|)$ is a decreasing function of $\|\bm h\|$.
\end{teo}

Example \ref{ex:matern1}, therefore, is a direct consequence of Theorems \ref{th:main} and \ref{th:monotonia_matern}. Moreover, in Theorem \ref{th:monotonia_matern} it is not necessary to restrict the smoothness parameter of bivariate Mat\'{e}rn covariance model, $\nu_{XY}=m+0.5,~m\in\mathds{N}_0$, as in Example \ref{ex:matern1}.

The Generalized Wendland family of covariance functions \citep{Gneiting:2002} is defined, for an integer $\kappa>0$, as
\begin{equation}\label{eq:wendland}
    \mathcal{GW}(h;\kappa,\mu)=\left\{
    \begin{array}{ll}
       \dfrac{1}{B(2\kappa,\mu+1)}\displaystyle\int_h^{1}u(u^2-h^2)^{\kappa-1}(1-u)^{\mu}du, & 0\leq h <1,\\
        0, & h\geq1,
    \end{array}\right.
\end{equation}
where $B$ denotes the beta function and $\mu$ must be positive with a lower bound as given in \cite{Gneiting:2002}. By continuity, for $\kappa=0$, we have
\begin{equation*}
    \mathcal{GW}(h;0,\mu)=\left\{
    \begin{array}{ll}
       (1-h^2)^{\mu}, & 0\leq h <1,\\
        0, & h\geq1.
    \end{array}\right.
\end{equation*}
Specific conditions for $\mu$ can also be found in \cite{Bevilacqua:2019}.
\begin{lem}\label{lemma:wendland}
The  function in Equation \eqref{eq:wendland} is decreasing in $h$ for all $\kappa\geq0$.
\end{lem}

For a bivariate Gaussian random field the Wendland-Gneiting covariance function has been extended \citep{Daley:2015} and is defined as
{\small
 \begin{align}
C_X(h)&=\sigma_X^2 c_{11}b_{11}^{\nu+2\kappa+1}B(\nu+2\kappa-1,\gamma_{11}+1)\mathcal{GW}\left(\dfrac{h}{b_{11}};\kappa,\nu+\gamma_{11}+1\right), \label{eq:wend1}\\
C_Y(h)&=\sigma_Y^2 c_{22}b_{22}^{\nu+2\kappa+1}B(\nu+2\kappa-1,\gamma_{22}+1)\mathcal{GW}\left(\dfrac{h}{b_{22}};\kappa,\nu+\gamma_{22}+1\right), \label{eq:wend2}\\
C_{XY}(h)&=\rho_{XY}\sigma_X \sigma_Y c_{12}b_{12}^{\nu+2\kappa+1}B(\nu+2\kappa-1,\gamma_{12}+1)\mathcal{GW}\left(\dfrac{h}{b_{12}};\kappa,\nu+\gamma_{12}+1\right),\label{eq:wend3}
\end{align}}
where $\sigma_X^2>0, \sigma_Y^2>0,$ and $\rho_{XY}$ is the co-located correlation coefficient between $X(\cdot)$ and $Y(\cdot)$. Specific conditions for the parameters $c_{ii},$ $b_{ii}$, $\gamma_{ii}, i=1,2,$ and $b_{12}$, $c_{12}$, $\gamma_{12}$, $\nu$, and $\kappa$ are needed so that the covariance model described in equations \eqref{eq:wend1}--\eqref{eq:wend3} is positive definite \citep{Daley:2015}. In this model we have that
{\small
\begin{equation}\label{eq:cov_wend}
\sigma_D^2(h)=\sigma_X^2+\sigma_Y^2-2\rho_{XY}\sigma_X \sigma_Y c_{12}b_{12}^{\nu+2\kappa+1}B(\nu+2\kappa-1,\gamma_{12}+1)\mathcal{GW}\left(\dfrac{h}{b_{12}};\kappa,\nu+\gamma_{12}+1\right).
\end{equation}}
\begin{teo}\label{th:monotonia_wendland}
Suppose that $\sigma_D^2(h)$ is as in Equation \eqref{eq:cov_wend}, and assume that $\rho_{XY}\geq0$ and $c_{12}\geq0$. Then, $\sigma_D(h)$ is an increasing function of $h$. In consequence, the conditions of Theorem \ref{th:main} are satisfied and  $\psi_c(\|\bm h\|)$ is a decreasing function of $\|\bm h\|$.
\end{teo}

\section{Probability of agreement for spatiotemporal processes}\label{sec:saptiotemporal}

Assume that $Z(\bm s, t)$, $\bm s\in D\subset\mathds{R}^2$,  $t\in \mathds{Z}_0^{+}$ is a stationary Gaussian spatiotemporal process with mean $0$ and covariance function $C(\bm h, u)=\text{Cov}(Z(\bm s, t), Z(\bm s+\bm h, t+u)$. Let
$$Y(\bm s, t)=\mu(\bm s, t)+Z(\bm s, t),$$ where the mean function $\mu(\bm s, t)=\bm F(\bm s, t)\bm\beta$, with $\bm F(\bm s, t)$ known and $\bm\beta$ is a vector of unknown parameters.

Now, let us define the difference
\begin{equation}\label{eq:D_st}
D(\bm s, t, \bm h, u)=Y(\bm s, t)-Y(\bm s+\bm h, t+u).
\end{equation}
The quantity defined in Equation \eqref{eq:D_st} measures  the discrepancy between the process and itself for a spatial separation $\bm h$ and temporal separation $u$. Then, under the Gaussian assumption
$$D(\bm s, t, \bm h, u)\sim \mathcal{N}(\mu_D(\bm h, u), \sigma^2_D(\bm h, u)),$$
where
\begin{eqnarray*}
 \mu_D(\bm h, u)&=&[\bm F(\bm s+\bm h, t+u)-\bm F(\bm s, t)]\bm\beta,\\[4pt]
 \sigma^2_D(\bm h, u)&=&2C(\bm 0, 0)-2C(\bm h, u):=2\gamma(\bm h, u),
\end{eqnarray*}
and $\gamma(\bm h, u)$ is the spatiotemporal semivariogram \citep{Sherman:2011}.
Consequently, a natural extension of the probability of agreement between  $Y(\bm s, t)$ and $Y(\bm s+\bm h, t+u)$ is
\begin{equation}\label{eq:spat_time_concordance}
\psi_c(\bm h, u)=\text{P}[\ |D(\bm s, t, \bm h, u)|\leq c], \ \ c>0,
\end{equation}
Therefore, the PA takes the form
\begin{equation}\label{eq:psi_c2}
\psi_c(\bm h, u)= \Phi\left(\frac{c-\mu_D(\bm h, u)}{\sigma_D(\bm h, u)}\right)- \Phi\left(-\frac{c+\mu_D(\bm h, u)}{\sigma_D(\bm h, u)}\right),
\end{equation}
where $\Phi(\cdot)$ is as in Equation \eqref{eq:concor2}.
As an illustration, if $\mu(\bm s, t)=\beta_0+\beta_1t$ (linear trend in  time) and $C(\bm h, u)=\sigma^2\exp(-\|\bm h\|/\phi_s)\exp(-| u|/\phi_t)$ (separable exponential covariance model) the mean and variance are
$$\mu_D(\bm h, u)=\beta_1u,\quad\text{and}\quad\sigma^2_D(\bm h, u)=2\sigma^2\left[1-\exp\left(-\dfrac{\|\bm h\|}{\phi_s}\right)\exp\left(-\dfrac{| u|}{\phi_t}\right)\right].$$
Thus, Equation \eqref{eq:psi_c2} can be written as
\begin{equation*}
\psi_c(\bm h, u)= \Phi\left(\frac{c-u\beta_1}{\sigma_D(\bm h, u)}\right)- \Phi\left(-\frac{c+u\beta_1}{\sigma_D(\bm h, u)}\right).
\end{equation*}
No matter the sign of $\beta_1$,  $\psi_c(\bm h, u)$ decreases as $u$ increases, which is in agreement with the fact that the PA become smaller when  separation over time is enlarged.

Note that we can also use Theorems \ref{th:main}-\ref{th:monotonia_wendland} and Equation \eqref{eq:psi_c2} to consider two different spatial processes in time. In this case, the covariance function has the same structure, but $C_X(\bm{h})=C_Y(\bm{h})$. Indeed, by replacing $\sigma_D(\|\bm{h}\|)$ by $\sigma_D(h,u)$ in Theorem \ref{th:main}, we obtain the same result if $\sigma_D(\|\bm{h}\|,u)$ is a decreasing function of $\|\bm{h}\|$ for fixed $u$. Moreover, for fixed $\|\bm h\|$, if $\sigma_D(\|\bm h\|,u)$ is an increasing function of $u$ then $\psi_c(\|\bm h\|,u)$ is a decreasing function of $u$. The proof is virtually identical to that of Theorem \ref{th:main}.

\section{Estimation}\label{sec:estimation}

The purpose of this section is to describe the estimation of the PA defined in Equation \eqref{eq:psi_c}. Here we emphasize that the variance of the difference depends on the parameters of the correlation structure, which we denote as
  $\sigma_D(\bm{h})=\sigma_D(\bm{h},\bm\theta)$, where $\bm\theta \in \mathbb{R}^q,~ q\in \mathbb{N}$, is a parameter vector associated with  the covariance function.
The next definition stresses the dependence of the PA on $\bm \theta$.

\begin{deft}
Suppose that $(X(\bm s), Y(\bm s))^{\top}$ is a bivariate second-order stationary random field with $\bm s, \bm h \in \mathbb{R}^2$, mean $(\mu_X, \mu_Y)^{\top}$, and parametric covariance function
$\bm C(\bm h;\theta)$. The probability
of agreement between processes $X(\bm s)$ and $Y(\bm s+\bm h)$ is defined through
\begin{equation*}
 \psi_c(\bm h;\mu_D,\bm\theta)= \Phi\left(\frac{c-\mu_D}{\sigma_D(\bm h;\bm\theta)}\right)-\Phi\left(-\frac{c+\mu_D}{\sigma_D(\bm h;\bm\theta)}\right),
\end{equation*}
where $\mu_D=\mu_X-\mu_Y$ and  $\sigma^2_D(\bm h;\bm\theta)=C_X(\bm 0;\bm\theta)+C_Y(\bm 0;\bm\theta )-2C_{XY}(\bm h;\bm\theta).$
\end{deft}
\noindent If $\widehat{\mu}_D$, and $\widehat{\bm\theta}$ are estimators of $\mu_D$, and $\bm\theta$, respectively, obtained from the sample $(X(\bm s_1), Y(\bm s_1))^{\top}$, $\ldots, (X(\bm s_n), Y(\bm s_n))^{\top}$, then the plug-in estimator of $\psi_c(\bm h;\mu_D,\bm\theta)$ is denoted as $\widehat{\psi}_c(\bm h)=\psi_c(\bm h;\widehat{\mu}_D,\widehat{\bm\theta})$.

\begin{lem}\label{lem:desv}
If $\bm V^{-1/2}_{\bm\theta}(\widehat{\bm\theta}-\bm\theta)$ is consistent and $\bm V^{-1/2}_{\bm\theta}(\widehat{\bm\theta}-\bm\theta) \overset{\mathcal{D}}{\longrightarrow} \mathcal{N}_q(\bm0,\bm{\mathbb{I}_q})$ as $n\to \infty$, then $$\widehat{\sigma}_D(\bm h;\bm\theta)=\sqrt{C_X(\bm 0;\widehat{\bm\theta})+C_Y(\bm 0;\widehat{\bm\theta} )-2C_{XY}(\bm h;\widehat{\bm\theta})}$$  is consistent and asymptotically Gaussian with
\begin{eqnarray}\nonumber
 \mathbb{E}[\widehat{\sigma}_D(\bm h;\bm\theta)] &\approx& \sqrt{C_X(\bm 0;\bm\theta)+C_Y(\bm 0;\bm\theta )-2C_{XY}(\bm h;\bm\theta)},\\[4pt] \label{eq:var_sigD}
 \mathrm{var}[\widehat{\sigma}_D(\bm h;\bm\theta)] &\approx& \dfrac{1}{4\sigma^2_D(\bm h;\bm\theta)} \nabla \sigma^2_D(\bm h;\bm\theta)^{\top} \bm V_{\bm\theta}\nabla \sigma^2_D(\bm h;\bm\theta).
\end{eqnarray}
\end{lem}

\begin{teo}\label{teo4}
Suppose that $(X(\bm s), Y(\bm s))^{\top}$ is a bivariate stationary Gaussian  random field. Assume that $\widehat{\mu}_D$ is consistent, $(\widehat{\mu}_D-\mu_D)/\sqrt{V_{\mu_D}(\bm\theta)} \overset{\mathcal{D}}{\longrightarrow} \mathcal{N}(0,1)$, $\widehat{\bm\theta}$ is consistent, $\bm V^{-1/2}_{\bm\theta}(\widehat{\bm\theta}-\bm\theta) \overset{\mathcal{D}}{\longrightarrow} \mathcal{N}_q(\bm0,\bm{\mathbb{I}}_q)$, and $\widehat{\mu}_D$ is independent of $\widehat{\bm\theta}$. Then,   $\widehat{\psi}_c(\bm h)$ is consistent and asymptotically Gaussian with
\begin{eqnarray*}
 \mathbb{E}[\widehat{\psi}_c(\bm h)]&\approx& \psi_c(\bm h;\mu_D,\bm\theta)
 \\[2pt]
 \mathrm{var}[\widehat{\psi}_c(\bm h)] &\approx& \dfrac{2}{\pi}\exp\left\{-\dfrac{(c-\mu_D)^2}{\sigma^2_D(\bm h;\bm\theta)}\right\}\left[ V_{\mu_D}+\dfrac{(c-\mu_D)^2}{\sigma^2_D(\bm h;\bm\theta)} V_{\sigma_D}\right].
\end{eqnarray*}
\end{teo}

As a consequence of the limiting distribution established in Theorem
\ref{teo4}, an approximate hypothesis test for the PA can be constructed. Consider the
null hypothesis
$$\text{H}_0:\psi_c(\|\bm h\|, \mu_D,\bm \theta)=\psi_c^{(0)},\ \  0\leq \psi_c^{(0)}\leq 1,$$
versus one of the following three alternative hypotheses $\text{H}_1:\psi_c(\|\bm h\|, \mu_D,\bm \theta) \neq \psi_c^{(0)}$, $\text{H}_1:\psi_c(\|\bm h\|, \mu_D,\bm \theta)> \psi_c^{(0)}$, or $\text{H}_1:\psi_c(\|\bm h\|, \mu_D,\bm \theta)<\psi_c^{(0)}.$ When  $\psi_c^{(0)}= 0.95,$ this hypothesis test is relevant because it compares the PA with the nominal value suggested by \cite{Stevens:2017b} for a fixed $\bm h$. In practice, if under the conditions of Theorem \ref{th:main}, the test of

$$\text{H}_0:\psi_c(0, \mu_D,\bm \theta)=0.95 \ \text{versus} \ \text{H}_1 :\psi_c(0, \mu_D,\bm \theta) <0.95$$
can be considered; if $\text{H}_0$ is rejected, then  $\text{H}_0:\psi_c(\|\bm h\|, \mu_D,\bm \theta)=0.95$ is rejected for all $\|\bm h\|$, because of the monotone property of the PA.

 In a spatiotemporal context, denote the covariance function parameterized by $\bm\theta$ as $C(\bm h, u;\bm\theta)$. Then, the PA in this case is
 $\psi_c(\bm h, u; \bm\beta,\bm\theta)$, similar to that defined in Equation \eqref{eq:psi_c2}, with $\sigma_D(\bm h,u)=\sigma_D(\bm h,u;\bm\theta)$. The plug-in estimator of $\psi_c(\bm h, u; \bm\beta,\bm\theta)$ is $\widehat{\psi}_c(\bm h, u)=\psi_c(\bm h, u; \widehat{\bm\beta},\widehat{\bm\theta})$. Thus, if $(\widehat{\bm\beta},\widehat{\bm\theta})$ is a consistent estimator of $(\bm\beta,\bm\theta)$, Theorem \ref{teo4} applies to $\widehat{\psi}_c(\bm h, u)$ considering $X(\bm s)=Y(\bm s, t)$ and $Y(\bm s)=Y(\bm s, t+u)$ for a fixed  $u$.

\section{Numerical Experiments}\label{sec:numerical_exp}
We carried out two numerical experiments to gain more insights into the properties of the PA for finite sample sizes. The first was a sensitivity analysis that examined how variation in key parameters affected the estimation of the PA for Gaussian random fields with a specific covariance function. The second was a Monte Carlo simulation study that considered spatiotemporal processes with a linear trend and either separable or non-separable covariance structures. In all cases, the estimates were obtained using a pairwise maximum-likelihood method implemented in the R software system version 4.0.5 \citep{R:2022}. Code is available at \url{https://github.com/JAcosta-Hub/Comparing-two-spatial-variables-with-the-probability-of-agreement}.

\subsection{Bivariate Gaussian random field}
\begin{figure}[hpt]
\centering
\includegraphics[scale=0.5]{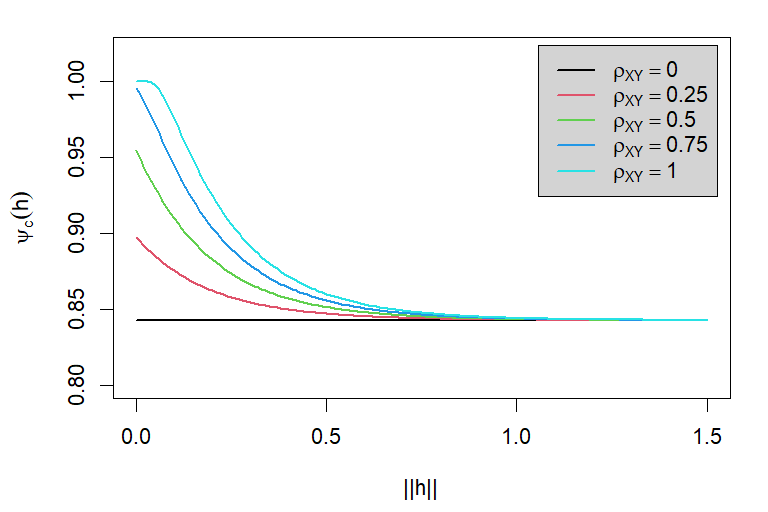}\hspace{2mm}
\includegraphics[scale=0.5]{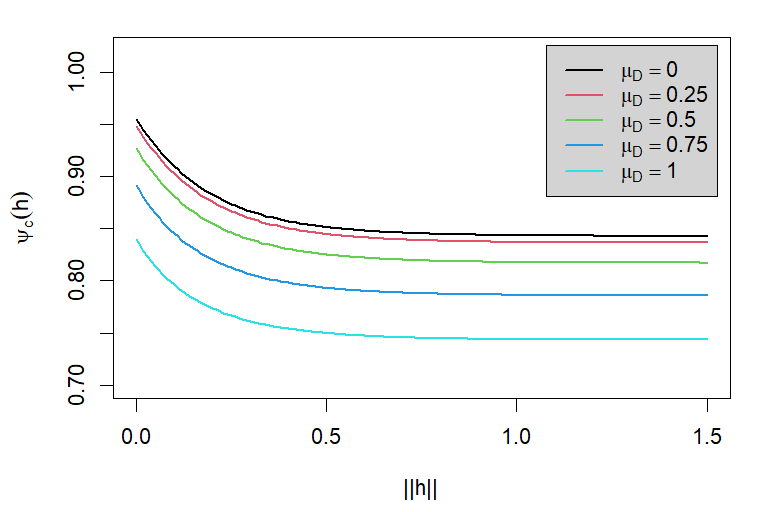}\\
\includegraphics[scale=0.5]{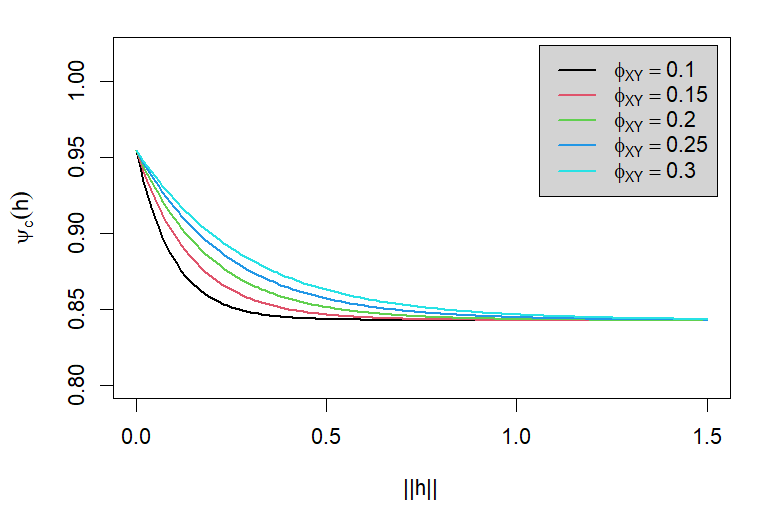}\hspace{2mm}
\includegraphics[scale=0.5]{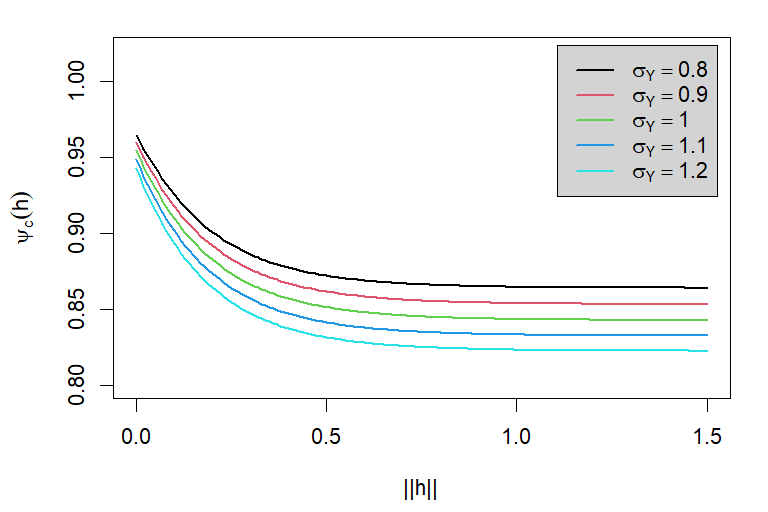}
\caption{The effect of variation in $\rho_{XY}$ (top left), $\mu_D$ (top right), $\phi_{XY}$ (bottom left), and $\sigma_Y$ (bottom right), on the behavior of PA as a function of lag $\|h\|$ in a bivariate Gaussian random field with a Mat\'{e}rn covariance structure. Note differences in range limits of the \textit{y}-axis among the four panels.}
\label{fig:matern}
\end{figure}

Let $\bm Z(\bm s)=(X(\bm s), Y)\bm s))^{\top}$ be a bivariate  stationary Gaussian random field with a Mat\'{e}rn covariance function as in equations \eqref{eq:mat1}--\eqref{eq:mat3}, where $\sigma_X^2=1$, $\nu_X=\nu_Y=\nu_{XY}=0.5$, $a_X=a_Y=1$, $a_{XY}\in\{0.1, 0.15, 0.2, 0.25, 0.3\}$, $\mu_X=1$, $\mu_Y\in\{0,0.25,0.5,0.75,1\}$, $\rho_{XY}\in\{0,0.25,0.5,0.75,1\}$, and $\sigma^2_{Y}\in\{0.8,0.9,1,1.1,1.2\}$. Assuming that $c=1$ and a fixed covariance parameter, we examined the behavior of the PA, $\psi_c(\|\bm{h}\|)$ (Fig. \ref{fig:matern}). In all cases, we observed that $\psi_c(\|\bm{h}\|)$ is a decreasing function of $\|\bm{h}\|$ and for large $\|\bm{h}\|$, reaches a fixed value corresponding to the uncorrelated case. As expected, for a fixed $\|\bm{h}\|$, PA increases with the correlation in the data ($\rho_{XY}$). Finally, as either $\mu_D$ or $\sigma_D$ increases, PA decreases.

\subsection{Gaussian spatiotemporal process}

The Monte Carlo simulation study considered a spatiotemporal process as defined in \S\ref{sec:saptiotemporal}. For the purposes of this study, we assumed that
\begin{equation*}
    \mu(\bm s, t)=a_0+a_1 t,
\end{equation*}
and for the correlation structure, we considered both separable and non-separable cases:
\begin{eqnarray*}
 \text{(Exponential-Separable)}&\quad& R(\bm h, u)=\exp\left(-\dfrac{\|\bm h\|}{\phi_s}\right)\exp\left(-\dfrac{|u|}{\phi_t}\right)\\[3pt]
 \text{(Iacocesare-Non-separable)} &\quad& R(\bm h, u)=\left(1+\left(\dfrac{\|\bm h\|}{\phi_s}\right)^{\alpha_s}+\left(\dfrac{|u|}{\phi_t}\right)^{\alpha_t}\right)^{-\beta}
\end{eqnarray*}
In the absence of a nugget, the covariance function is $C(\bm h, u)=\sigma^2R(\bm h,u)$.

We used the \textit{GeoModels} library version 1.0.0 \citep{GM:2022} in R for the simulations, as it allowed us to simulate spatiotemporal processes with linear mean and our defined correlation structures. A regular grid of size $N_S\times N_S$ was considered for the spatial coordinates, $N_T$ points in time from 1 to $N_T$ with step 1; examples are shown in Figs. 2 and 3 of the Supplementary Material.



The estimates of the spatial scale (extent) parameter $\phi_s$ in the covariance terms had very large ranges, extending from $<0$ to greater than the maximum size of the grid. Thus, we only estimated PA for values of $\phi_s$ such that $0 < \phi_s <$ the maximum size of the grid (in this case, 50).

\begin{figure}[hpt]
\centering
\includegraphics[scale=0.82]{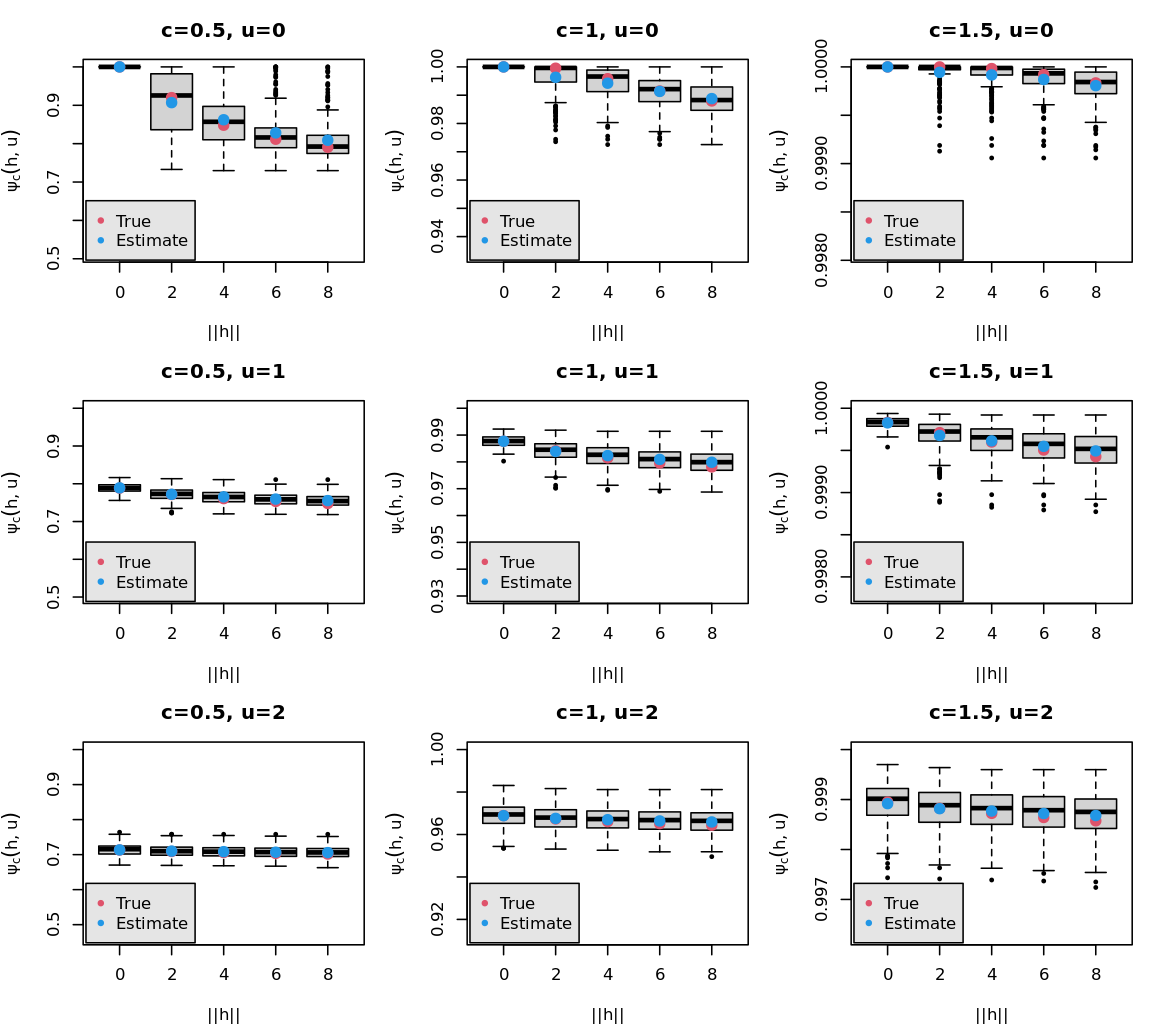}
\caption{Estimates of the probability of agreement as a function of $\|\bm h\|$, $u$ and $c$ for a spatiotemporal Gaussian process with a positive linear trend and a non-separable Iacocesare covariance structure with fixed parameters given in Table \ref{tab:spatiotemp}, and for $N_s=50$. Note differences in range limits of the \textit{y}-axis among the nine panels.}
\label{fig:psi_st_nonseparable2}
\end{figure}

The parameter estimates for the spatiotemporal process with a separable covariance structure (Fig. 2 in the Supplementary Material) and two different values for the trend parameter $a_1$ ($-0.1$ and $0.1$) are given in Table \ref{tab:spatiotemp}; the estimators were practically unbiased and consistent. The corresponding estimates of PA for different values of $h, u$ and $c$, and fixed parameters given in Table \ref{tab:spatiotemp} are shown in Figs. 4-7 of the Supplementary Material. The corresponding results from the simulations of a spatiotemporal process with a non-separable covariance structure and identical values for the trend parameter $a_1$ ($-0.1$ and $0.1$) are given, respectively, in Table \ref{tab:spatiotemp} and Figs. 8-10 of the Supplementary Material, and Table \ref{tab:spatiotemp} and Fig. \ref{fig:psi_st_nonseparable2}. As with the separable case, the estimators were practically unbiased and consistent. The model with the Iacocesare covariance had a slightly better performance than the model with the exponential-separable covariance and a higher percent of valid cases used for estimating the parameters.
\begin{table}[htp]
\centering
\caption{Parameter estimates for the spatiotemporal process with an exponential separable and a non-separable Iacocesare covariance structure. For the time trend, the cases of positive and negative slope are included. ``Percent valid'' is the percentage of simulations for which estimates of $\phi_s$ were greater than zero and less than the maximum size of the grid.}
\scalebox{0.875}{
\begin{tabular}{lcrrrrrrrrrc}
  \toprule
Covariance& & & & & & & & & & & Percent \\
Model&$(N_S, N_T)$& & $a_0$ & $a_1$ & $\phi_s$ & $\phi_t$ & $\sigma^2$ & $\alpha_s$ & $\alpha_t$ & $\beta$ &  valid \\
  \midrule
\multirow{10}{*}{Exponential} & &true & 0.500 & -0.100 & 6.676 & 1.000 & 0.100 &  &  &  & \\ \cmidrule{3-11}
&\multirow{2}{*}{$(20,10)$}&mean & 0.510 & -0.101 & 7.714 & 0.789 & 0.090 & &&& \multirow{2}{*}{$77.6\%$}  \\
&&sd & 0.145 & 0.023 & 3.342 & 0.231 & 0.014 & &&&\\    \cmidrule{2-12}
&\multirow{2}{*}{$(50,10)$}&mean & 0.504 & -0.101 & 7.944 & 0.952 & 0.097 &&&&\multirow{2}{*}{$74.0\%$}  \\
&&sd & 0.070 & 0.011 & 3.816 & 0.126 & 0.008 &&&& \\    \cmidrule{2-12}
&&true & 0.500 & 0.100 & 6.676 & 1.000 & 0.100 & &&&\\ \cmidrule{3-11}
&\multirow{2}{*}{$(20,10)$}&mean & 0.511 & 0.098 & 7.437 & 0.830 & 0.090 & &&&\multirow{2}{*}{$82.4\%$}  \\
&&sd & 0.145 & 0.024 & 2.974 & 0.226 & 0.014 & &&&\\  \cmidrule{2-12}
&\multirow{2}{*}{$(50,10)$}&mean & 0.506 & 0.099 & 7.548 & 0.940 & 0.096 & &&&\multirow{2}{*}{$84.4\%$}  \\
&&sd & 0.063 & 0.010 & 3.706 & 0.113 & 0.007 & &&& \\  \midrule
\multirow{10}{*}{Iacocesare} &  &true & 0.500 & -0.100 & 6.676 & 1.000 & 0.100 & 1.000 & 1.000 & 2.000 & \\ \cmidrule{3-11}
&\multirow{2}{*}{$(20,10)$}& mean & 0.513 & -0.102 & 7.250 & 0.987 & 0.091 & 1.891 & 2.180 & 2.534 & \multirow{2}{*}{$62.0\%$} \\
&&sd & 0.132 & 0.019 & 2.475 & 0.887 & 0.009 & 2.246 & 2.106 & 1.730 & \\  \cmidrule{2-12}
&\multirow{2}{*}{$(50,10)$}&mean & 0.506 & -0.101 & 7.092 & 1.104 & 0.096 & 1.698 & 1.358 & 2.268 &\multirow{2}{*}{$85.6\%$}  \\
&&sd & 0.086 & 0.011 & 2.407 & 0.750 & 0.005 & 1.738 & 0.895 & 1.209 & \\    \cmidrule{2-12}
&&true & 0.500 & 0.100 & 6.676 & 1.000 & 0.100 & 1.000 & 1.000 & 2.000 & \\ \cmidrule{3-11}
&\multirow{2}{*}{$(20,10)$}& mean & 0.493 & 0.099 & 6.896 & 0.979 & 0.090 & 1.485 & 2.413 & 2.881 & \multirow{2}{*}{$67.2\%$} \\
&&sd & 0.132 & 0.020 & 2.272 & 0.641 & 0.009 & 2.009 & 2.173 & 2.299 & \\  \cmidrule{2-12}
&\multirow{2}{*}{$(50,10)$}&mean & 0.504 & 0.101 & 6.778 & 1.051 & 0.096 & 1.452 & 1.497 & 2.217 &\multirow{2}{*}{$82.8\%$}  \\
&&sd & 0.082 & 0.012 & 2.072 & 0.579 & 0.005 & 1.765 & 1.128 & 1.055 & \\
   \bottomrule
\end{tabular}
\label{tab:spatiotemp}
}
\end{table}




\section{An Empirical Example}\label{sec:application}

\subsection{Motivation}
Near-Earth remote sensing provides a great deal of information about ongoing environmental change and its effects on the Earth's climate \citep[e.g.,][]{Richardson:2007, Richardson:2018, Yang:2013}. Of particular interest are the times of year when trees in the northern hemisphere emerge from dormancy and produce new leaves (``spring green-up'') and when the leaves of these same trees senesce in the fall before the trees go dormant for the winter. The growing season is the time between the spring green-up and leaf senescence in the fall, and is that time of year when forests in the northern hemisphere remove a substantial amount of carbon dioxide from the atmosphere \citep{Barichivich:2012}. There is substantial evidence that because of ongoing, anthropogenically-driven climate change, on average spring green-up is occurring earlier in the year \citep[e.g.,][]{Richardson:2007, Keenan:2014} and leaf senescence is occurring later in the fall \citep[e.g.,][]{Moon:2022}.

\subsection{Imagery}

\begin{figure}[htp]
	\centering
	\includegraphics[width=.9\textwidth]{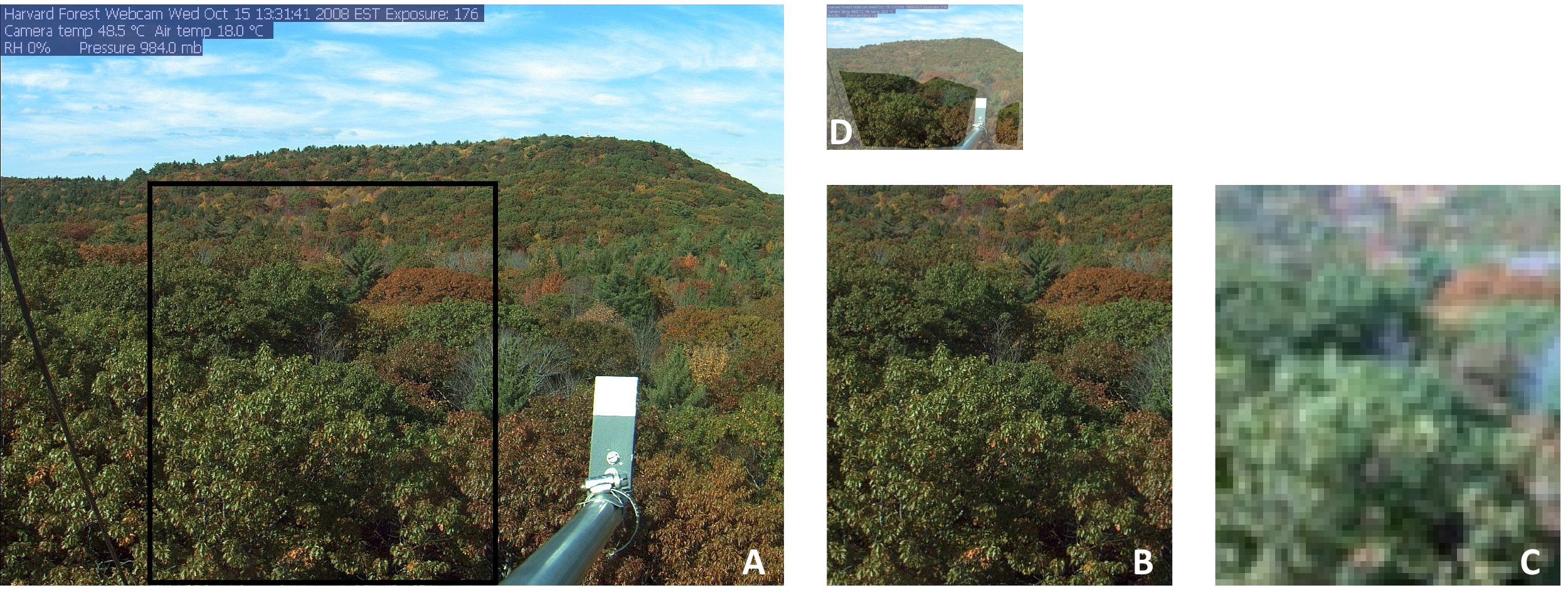}
	\caption{\textbf{A.} The PhenoCam image taken by a stationary camera from the EMS tower at Harvard Forest, Massachusetts, USA on 15 October 2008 at 13:31 (UTC $-4$). A rectangular section ($570 \times 660$ pixels) of the image was clipped  (\textbf{B}; black outline in \textbf{A}) and then down-scaled (to $44 \times 58)$ pixels (\textbf{C}) for estimation of the spatial PA. Note that our rectangular image is different from the clipped ``region of interest'' (ROI; unmasked area in \textbf{D}) analyzed by the PhenoCam network.}
	\label{fig:im_15.10.08}
\end{figure}

We analyzed a series of 15 annual images of the same scene taken in mid-October from 2008--2022 (Fig. \ref{fig:im_15.10.08}A). This is the time of year when leaves of deciduous trees are senescing, which leads to the spectacular display of fall colors in New England (USA), Japan, and other parts of the northern hemisphere. We chose images taken in a 3-day window (12--15 October) each year, as this represents the approximate historical ``peak'' of fall colors in New England. As the regional climate has warmed, however, this peak has begun to show a shift towards later dates, and identifying the rate and spatial patterning of this shift is of interest to ecologists, foresters, tourism boards, and economists \citep[e.g.,][]{Moon:2022}.

The 15 images we used were taken from the database of the PhenoCam Network,\footnote{\url{https://phenocam.nau.edu/}} a network of more than 700 fixed observation sites across North America and elsewhere in the world that since 2008 has collected high-frequency and high-resolution imagery with networked digital cameras to track the timing of vegetation change (phenology) in a range of ecosystems \citep{Seyednasrollah:2019}. We used images from the Harvard Forest, where they have been captured at 30--60-minute intervals since 2008 with a 2048 $\times$ 1636-pixel CMOS sensor in an outdoor StarDot NetCam XL 3MP camera \citep{Richardson:2021}.

To focus attention on the forest canopy, we clipped a $570 \times 660$-pixel rectangular section from each image Fig. \ref{fig:im_15.10.08}B). The edges of the clipped image (black rectangle in Fig. \ref{fig:im_15.10.08}A) were chosen to maximize the size of the clipped image while avoiding sky (top), wires (left) and other instrumentation on the extended boom (lower right). The high-resolution clipped image (Fig. \ref{fig:im_15.10.08}B) was then downscaled $\approx$100-fold (to $44 \times 58$ pixels; Fig. \ref{fig:im_15.10.08}C)) for further analysis and estimation of spatiotemporal PA. Downscaling was done by rasterizing the image using adjoining $15 \times 15$-pixel windows; we used the mean RGB value from these windows in the rasterized image (Fig. \ref{fig:im_15.10.08}C). This downscaling was done for two reasons. First, reasonable values of $\|h\|$ (i.e., $< 15$) would have been within a single leaf of the high-resolution image, and it is of more interest to look at changes among leaves and among entire trees. Second, we estimated that estimating the PA of the two high-resolution clipped images would require $\approx$ 6 PB of RAM, whereas the estimation of the downscaled images, which were close to the same size as those used for our simulation studies (Figs. 2 and 3 in the Supplementary Material), preserved sufficient visual differences among trees while being computationally more manageable.

Last, for each pixel, we calculated its green chromatic coordinate ($G_{cc}$), an index of ``greenness'' that captures the phenological stage of tree leaves, is associated with carbon flux from ecosystems, and is estimated from repeated images of forest canopies \citep{Richardson:2018}. For a pixel in an RGB image, $G_{cc}$ is calculated as $G_{cc}=\frac{G_{DN}}{G_{DN}+R_{DN}+B_{DN}}$, where $\cdot_{DN}$ is the digital number of the Green (G), Red (R), and Blue (B) channels, respectively, assigned to each pixel in a digital image. The PhenoCam network calculates $G_{cc}$ for each pixel and in their ``provisional'' data products reports its mean, and the 50\textsuperscript{th}, 75\textsuperscript{th}, and 90\textsuperscript{th} percentiles of the $G_{cc}$ for the ROI of each image, and their 1-day and 3-day running means and percentiles \citep{Richardson:2018}. Original, clipped, and rasterized images, and code used for rasterizing and analyzing these images are all available on GitHub \url{https://github.com/JAcosta-Hub/Comparing-two-spatial-variables-with-the-probability-of-agreement}.

\begin{figure}[ht!]
    \centering
    \includegraphics[width=.9\textwidth]{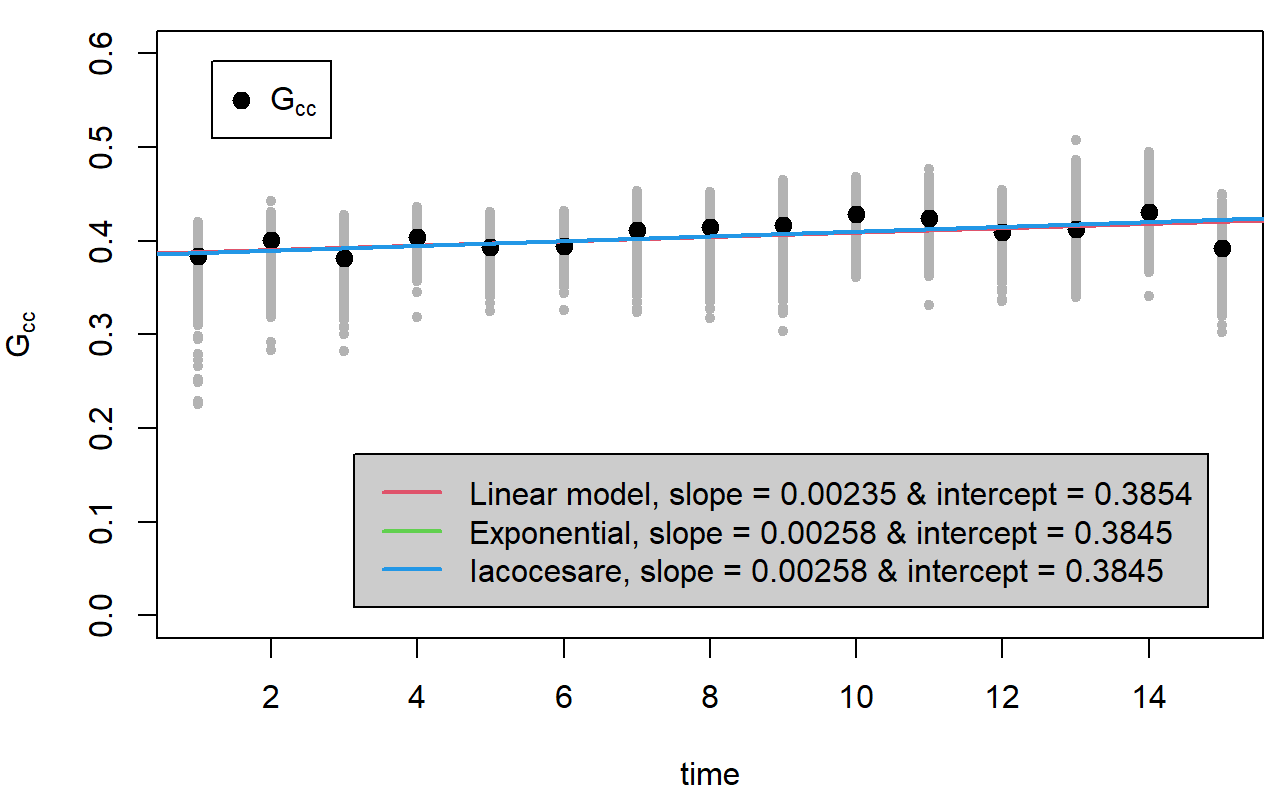}
    \caption{Green chromatic indices ($G_{cc}$) of each of the rasters in the 15 downscaled images (grey symbols) and the mean $G_{cc}$ estimated for the entire region of interest (ROI; masked area in Fig. \ref{fig:im_15.10.08}D) by the PhenoCam network. Also shown are the slopes and intercepts of the temporal trend in the $G_{cc}$ assuming independence (red line), separable (green), and non-separable (blue) spatial covariances.}
    \label{fig:GCCs}
\end{figure}

We note that our clipped rectangle is different in shape, but approximately the same size, as the ``region of interest'' (ROI) defined and analyzed by researchers who use these images for phenological studies (the unmasked area in Fig. \ref{fig:im_15.10.08}D). The ROI for each PhenoCam site is identified as the area of the image that maximizes the amount of vegetation of interest (i.e., deciduous forest at Harvard Forest) while avoiding sky, topographic features, instrumentation, other human artefacts (e.g., buildings, wires), and other areas of the image that could give seasonally biased results (e.g., soil covered by snow) \citep{Richardson:2007}. Our estimates of the mean $G_{cc}$ calculated by the PhenoCam network for the corresponding ROI fell within the range of $G_{cc}$ values for each pixel of our clipped and rasterized images (Fig. \ref{fig:GCCs}).

\subsection{Estimates}

The empirical variogram of the original data showed a strong temporal dependence, so we used linear regression independent of spatial covariance to remove the effect of a deterministic trend (slope = 0.00235, intercept = 0.3854; $P < 0.0001$ for both). Fig. 11 in the Supplementary Material shows the marginal empirical variograms after this trend had been removed (i.e., the empirical variogram of the residuals of the simple linear regression).


To model $G_{cc}$, we considered a linear temporal trend $\mu(\bm s, t)=\mu(t)=a_0+a_1 t$ and different covariance models (separable and non-separable), each with fixed nugget effect equal to 0.
 The value of the objective function (log composite likelihood) for the Exponential and Iacocesare models, respectively, were 19668351.95 and 19682852.64, and the values of the associated pseudo-Akaike information criterion were, respectively, -39336693.90 and -39365689.29.
 These results suggested a better fit to the data when using the Iacocesare non-separable covariance model. The parameter estimates (and the values we used to initialize the estimation routine) for the temporal trend and the covariance model are given in Table \ref{tab:estimated_iacocesare}.

The practical spatial range, defined as the distance at which 95\% of the sill (i.e., $\sigma^2$ in this case) is reached, will depend on the temporal separation. For the Iacocesare covariance model, this distance can be obtained by using the estimates of the parameters in this equation:

\begin{equation*}
    \text{Practical Range}=\phi_s\left(20^{1/\beta}-1-\left(\dfrac{|u|}{\phi_t}\right)^{\alpha_t}\right)^{1/\alpha_s},\qquad 20^{1/\beta}-1-\left(\dfrac{|u|}{\phi_t}\right)^{\alpha_t}>0.
\end{equation*}

Using the estimates given in  Table \ref{tab:estimated_iacocesare} and setting  $u=0$ (no time lag), the estimated practical range is $12.7$ pixels.


\begin{table}[htp]
\centering
\caption{Parameter estimates for the linear temporal trend and the Iacocesare covariance structure.}
\begin{tabular}{rrrrrrrrr}
  \toprule
 & $\widehat{a}_0$ & $\widehat{a}_1$ & $\widehat{\alpha}_s$ & $\widehat{\alpha}_t$ & $\widehat{\beta}$ & $\widehat{\phi}_s$ & $\widehat{\phi}_t$ & $\widehat{\sigma}^2$ \\
  \midrule
initial & 0.50000 & 0.01000 & 1.00000 & 1.00000 & 2.00000 & 6.67616 & 1.00000 & 0.00100 \\
  estimate & 0.38453 & 0.00258 & 1.06297 & 0.95371 & 1.94867 & 6.53170 & 2.08316 & 0.00044 \\
   \bottomrule
\end{tabular}
\label{tab:estimated_iacocesare}
\end{table}

\begin{figure}[hpt]
\centering
\includegraphics[width=.9\textwidth]{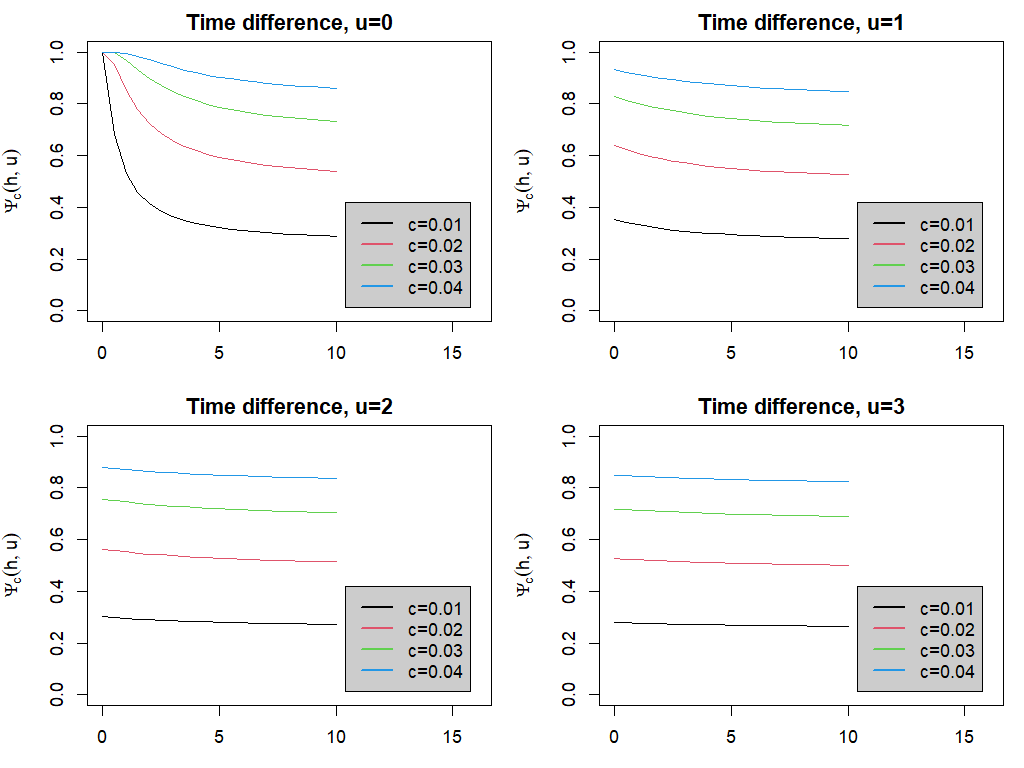}
\caption{Probability of Agreement as a function of $\|h\|$ for four time lags $u$ (years between images (individual panels), each with different of maximum acceptable differences $c$ (black, red, blue and green curves within each graph).}
\label{fig:PA_st}
\end{figure}

Finally, Fig. \ref{fig:PA_st} illustrates estimates of the PA of the $G_{cc}$ between images as a function of spatial lag $\|h\|$ for four different time lags ($u$) and four different maximum acceptable differences $c$. Regardless of the values of $u$ and $c$, PA decreases with increasing $\|h\|$, but the rate of decrease declines rapidly with increasing $u$. For $u>1$, PA is practically independent of $\|h\|$ (and $c$), which we interpret to mean that the spatiotemporal processes separated by two or more years are independent and are essentially Markovian in time.

\section{Discussion and future work}\label{sec:conclusions}

The probability of agreement has been generalized for the analysis of concordance between two georeferenced variables. This extension possesses monotonic properties---the PA declines as a function of the norm of the spatial lag---and thus quantifies the effective (or practical) spatial range. Our spatial PA is meaningful for isotropic processes. The hypothesis testing developed in Section \ref{sec:estimation} allows the estimation of the PA as a function of the spatial range, and provides a way to rule out spatial agreement when the null hypothesis is rejected for all $\|h\|$. Our theoretical extension of the PA also works for spatiotemporal processes with a trend, allowing for the analysis of nonstationary processes in the mean.

The monotonic properties of the PA for finite sample size were supported by Monte Carlo simulation experiments, which also showed that the parameter estimates using the composite likelihood have small bias and low variance. The value of $c$ plays a crucial role in the estimation and the PA is sensitive to the choice of it. In practice, the value of $c$ needs to be scaled by the square root of the sill in order to account for the  scale of the data.

The application presented in Section \ref{sec:application} illustrated that the PA can describe the change of a spatiotemporal variable in time while accounting for spatial and temporal dependence. In this particular case, the PA also provided information about the dependence of the trend on the spatial information of the past realizations of the process; such information may be of value in modeling the trend and developing such models should be a focus of future work. Although parameter estimates were not very sensitive to the choice of the covariance function, identification of the best covariance model (e.g., separable or non-separable, and types of each) can still be improved. If the trend cannot be modeled easily with a well-known function, prior exploration will be needed to characterize a parametric function that accurately identifies observed patterns. Nonparametric trend estimation \citep[e.g.,][]{Strandberg:2019} could also be used.

The computational efficiency of the composite likelihood method used in Sections \ref{sec:numerical_exp} and \ref{sec:application} deserves attention. The computational implementations used in this article worked well for images of the order of size $50\times 50$; we had to rasterize and downscale the original images used in Section \ref{sec:application} by two orders of magnitude before we could estimate the relevant parameters. Overcoming memory limitations to enable parameter estimation from much larger images (more than $10^6$ pixels) will require innovative parallelizable algorithms.

We also note that the Monte Carlo simulations and the application were developed and illustrated using spatial processes (images) defined on regular grids. However, there is no apparent reason that the proposed methods could not also be applied to irregularly spaced spatial data.

Finally, a natural but unexplored extension of the present work is the definition of the PA for spatiotemporal marked point processes. Because the randomness in point processes is in the location (as well as in the marks) and repeated observations of a given point process will yield a new set of locations, it is of theoretical interest to estimate the PA of point patterns that evolve through time. Such estimation would have immediate applicability to ecological spatial datasets, which are predominantly samples of point processes, not rasters \citep[e.g.,][]{Plant:2019}

\section*{Acknowledgements}
This work has been partially supported by the AC3E, UTFSM, under grant FB-0008, and from USM PI-L-18-20.
R. Vallejos also acknowledges financial support from CONICYT through the MATH-AMSUD program, grant  20-MATH-03. A. M. Ellison's work on this project was supported by a Fulbright Specialist Grant, Fulbright Chile, and the Universidad T\'{e}cnica Federico Santa Maria. M. de Castro's work was partially funded by CNPq, Brazil.


\appendix
\section*{Appendix}

{\bf Proof of Theorem \ref{th:main}}

Without loss of generality, we assume that the Gaussian process has mean $\bm 0$ and $h=\|\bm h\|$ in \eqref{eq:psi_c}.
First notice that
\begin{align*}
\psi_c(h)&=\Phi\left(\frac{c}{\sigma_D(h)}\right)-\Phi\left(-\frac{c}{\sigma_D(h)}\right)
=2\Phi\left(\frac{c}{\sigma_D(h)}\right)-1.
\end{align*}
Now, lets $h_1,h_2\in\mathbb{R}^{+}$ such that $h_1\leq h_2$, by hypothesis $\sigma_D(h_1)\leq\sigma_D(h_2)$, then
\begin{equation*}
    \dfrac{c}{\sigma_D(h_2)}\leq\dfrac{c}{\sigma_D(h_1)}
\end{equation*}
because $c>0$. Finally, as $\Phi$ is an increasing function, then
\begin{equation*}
   \Phi\left(\dfrac{c}{\sigma_D(h_2)}\right)\leq\Phi\left(\dfrac{c}{\sigma_D(h_1)}\right)\quad\Longleftrightarrow\quad 2\Phi\left(\dfrac{c}{\sigma_D(h_2)}\right)-1\leq2\Phi\left(\dfrac{c}{\sigma_D(h_1)}\right)-1.
\end{equation*}
Therefore, $\psi_c(h_2)\leq\psi_c(h_1)$ for all $h_1\leq h_2$, thus the proof is completed.
 \hfill $\square$

\bigskip
\noindent {\bf Proof of Theorem \ref{th:monotonia_matern}}
\smallskip

Note that
\begin{equation*}
    \sigma^{\prime}_D(h) = -\dfrac{\rho_{XY}\sigma_X \sigma_YM^{\prime}(h,\nu_{XY},a_{XY})}{\sigma_D(h)}.
\end{equation*}
Also note that  $\rho_{XY}\geq0$, $\sigma_X>0$, $\sigma_Y>0$ and $\sigma_D(h)>0$ for all $h>0$. Thus $\sigma^{\prime}_D(h)>0$ if and only if $M^{\prime}(h,\nu_{XY},a_{XY})<0$.
Without loss of generality, we assume that $a=1$ and $\nu=\nu_{XY}$ in \eqref{eq:mat}.
Noticing that the terms $M^{\prime}(h,\nu,1)$ and $g_{\nu}^{\prime}(h)$ have the same sign, where $g_{\nu}(h)=h^\nu K_{\nu}(h)$, and using the properties of the modified Bessel functions of the second kind \citep[][p.110]{Lebedev:1965}, we have that
\begin{equation*}
    g_{\nu}^{\prime}(h)=-h^{\nu}K_{\nu-1}(h).
\end{equation*}
Since $K_{\alpha}(x)=K_{-\alpha}(x)$ \citep[][p.110]{Lebedev:1965}, $K_{\alpha}(x)>0,$ for all $x>0$ and $\alpha\in \mathbb{R}$ \citep[][p.136]{Lebedev:1965}, it follows that $g_{\nu}^{\prime}(h)<0$, and the proof is complete.
\hfill $\square$

\bigskip
\noindent {\bf Proof of  Lemma \ref{lemma:wendland}}
\smallskip

Let $h_1, h_2\in \mathbb{R}$, such that $0\leq h_1\leq h_2$. If $h_1\geq1$, then $\mathcal{GW}(h_1;\kappa,\mu)=\mathcal{GW}(h_2;\kappa,\mu)=0$ and $\mathcal{GW}(\cdot;\kappa,\mu)$ is a monotone function, if $h_1<1$ and $h_2\geq1$, then $\mathcal{GW}(h_1;\kappa,\mu)\geq 0$, and  $\mathcal{GW}(h_2;\kappa,\mu)=0$, then $\mathcal{GW}(\cdot;\kappa,\mu)$ is a decreasing monotone function. If $h_2<1$, we distinguish the following two cases:
\begin{itemize}
\item  For $\kappa=0$, note that $0\leq 1- h_2^2\leq 1-h_1^2<1$ and $0\leq (1- h_2^2)^{\mu}\leq (1-h_1^2)^{\mu}<1$, therefore $\mathcal{GW}(h_1;\kappa,\mu)\geq\mathcal{GW}(h_2;\kappa,\mu)$.
\item For $\kappa\geq1$, we define $g(u,h)=u(u^2-h^2)^{\kappa-1}(1-u)^{\mu}/B(2\kappa,\mu+1)$. Clearly $g(u,h)\geq0$ for $0\leq h< u<1$, then $G(h)=\int_{h}^{1}g(u,h)du$ corresponds to $\mathcal{GW}(h;\kappa,\mu)$. Hence, by Leibniz's formulae,
\begin{equation*}
 G^{\prime}(h)=\int_{h}^{1} \dfrac{\partial g(u,h)}{\partial h}du-g(h,h) = -2h(k-1)\int_{h}^{1}\dfrac{g(u,h)}{u^2-h^2}du.
\end{equation*}
Because $h>0$,  $G^{\prime}(h)<0$ if and only if $\kappa>1$. When $\kappa=1$, the function  $g(u,h)=\tilde{g}(u)\geq0$, and $G^{\prime}(h)=-\tilde{g}(h)\leq0$.
\end{itemize}
Therefore $\mathcal{GW}(\cdot;\kappa,\mu)$ is a decreasing monotone function for $h_2<1$.  \hfill $\square$

\bigskip
\noindent {\bf Proof of  Theorem \ref{th:monotonia_wendland}}
\smallskip

Without  loss  of  generality,  we  assume  $b_{12}=1$ and note that $\sigma_D^2(h)$ is an increasing function of $h$ if and only if $\mathcal{GW}(h;\kappa,\mu)$ is a decreasing function in $h$ for all $\kappa$. Therefore, the result holds by Lemma \ref{lemma:wendland}, since $\nu+\gamma_{12}+1>0$.  \hfill $\square$

\bigskip
\noindent {\bf Proof of  Lemma \ref{lem:desv}}
\smallskip

Let $\widehat{\sigma}_D(\bm h;\bm\theta)=g(\widehat{\bm\theta})$. Applying a Taylor expansion of order 1 for $g(\widehat{ \bm \theta})$ around $\bm\theta$, we have that
\begin{equation*}
  \widehat{\sigma}_D(\bm h;\bm\theta)\approx g(\bm\theta) + \nabla g(\bm\theta)^{\top}(\widehat{\bm\theta}-\bm\theta).
\end{equation*}
Then,
\begin{equation*}
    \mathbb{E}[ \widehat{\sigma}_D(\bm h;\bm\theta)]\approx g(\bm\theta),\quad\text{and}\quad \text{var}[ \widehat{\sigma}_D(\bm h;\bm\theta)]\approx \nabla g(\bm\theta)^{\top}\bm V_{\bm\theta}\nabla g(\bm\theta).
\end{equation*}
Now, because $g(\bm\theta)=\sqrt{\sigma^2_D(\bm h;\bm\theta)}$ then $\nabla g(\bm\theta)=\nabla\sigma^2_D(\bm h;\bm\theta)/(2g(\bm\theta))$, where the $i$-th element of $\nabla\sigma^2_D(\bm h;\bm\theta)$ is given by
\begin{equation*}
    \dfrac{\partial \sigma^2_D(\bm h;\bm\theta)}{\partial \theta_i} = \dfrac{\partial C_X(\bm 0;\bm\theta)}{\partial \theta_i} +\dfrac{\partial C_Y(\bm 0;\bm\theta )}{\partial \theta_i} - 2\dfrac{\partial C_{XY}(\bm h;\bm\theta)}{\partial \theta_i}.
\end{equation*} \hfill $\square$

\bigskip
\noindent {\bf Proof of  Theorem \ref{teo4}}
\smallskip

Denote $\sigma_D=\sigma_D(\bm h, \bm\theta)$, $\widehat{\sigma}_D=\sigma_D(\bm h, \widehat{\bm\theta})$, $\psi_c(\bm h;\mu_D,\bm\theta)=\psi_c(\bm h;\mu_D,\sigma_D)$, and $\psi_c(\bm h;\widehat{\mu}_D,\widehat{\bm\theta})=\psi_c(\bm h;\widehat{\mu}_D,\widehat{\sigma}_D)$. Let $V_{\sigma_D}(\bm h)=\text{var}[\sigma_D(\bm h;\widehat{\bm\theta})]$ given in Equation \eqref{eq:var_sigD}.  By Lemma \ref{lem:desv} we have that $\sigma_D(\bm h;\widehat{\bm\theta})$ is consistent, and $(\sigma_D(\bm h;\widehat{\bm\theta})-\sigma_D(\bm h;\bm\theta))/\sqrt{V_{\sigma_D}}\overset{\mathcal{D}}{\longrightarrow}\mathcal{N}(0,1)$.
Now, using  the Delta method (approximation of order 1), it follows that
\begin{equation}\label{eq:psi_delta}
  \psi_c(\bm h;\widehat{\mu}_D,\widehat{\sigma}_D)\approx\psi_c(\bm h;\mu_D,\sigma_D)+a(\widehat{\mu}_D-\mu_D)+b(\widehat{\sigma}_D-\sigma_D),
\end{equation}
where
\begin{eqnarray*}
a&=&\left.\dfrac{\partial \psi_c(\bm h;\widehat{\mu}_D,\widehat{\sigma}_D)}{\partial \widehat{\mu}_{D}}\right|_{(\mu_D,\sigma_D)}~=~-\dfrac{2}{2\pi}\exp\left\{-\dfrac{(c-\mu_D)^2}{2\sigma^2_D}\right\},\\[4pt]
b&=&\left.\dfrac{\partial \psi_c(\bm h;\widehat{\mu}_D,\widehat{\sigma}_D)}{\partial \widehat{\sigma}_{D}}\right|_{(\mu_D,\sigma_D)}~=~-\dfrac{2}{2\pi}\exp\left\{-\dfrac{(c-\mu_D)^2}{2\sigma^2_D}\right\}\left(\dfrac{(c-\mu_D)^2}{\sigma^2_D}\right).
\end{eqnarray*}
 Applying expected value and variance in both sides of Equation \eqref{eq:psi_delta}, the result follows.
\hfill $\square$
\clearpage

\setcounter{figure}{0}

\begin{center}
    \Large{\bf{Supplementary Material}}
\end{center}

\begin{figure}[ht!]
\begin{subfigure}{.5\textwidth}
  \centering
     \includegraphics[scale=0.3]{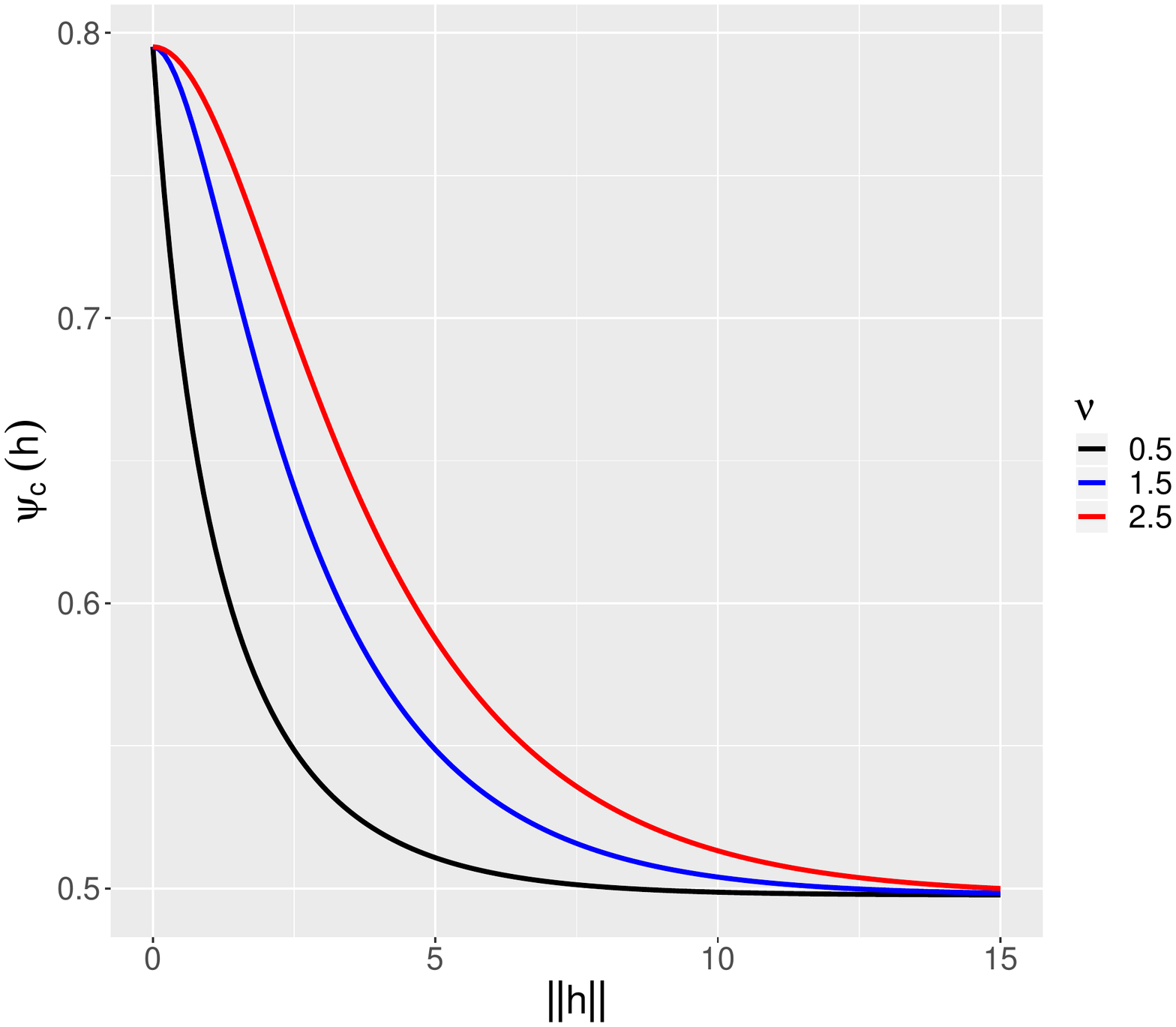}
\vspace{-10mm}
  \caption{}
\end{subfigure}
\vspace{-10mm}
\begin{subfigure}{.5\textwidth}
    \centering
    \includegraphics[scale=0.3]{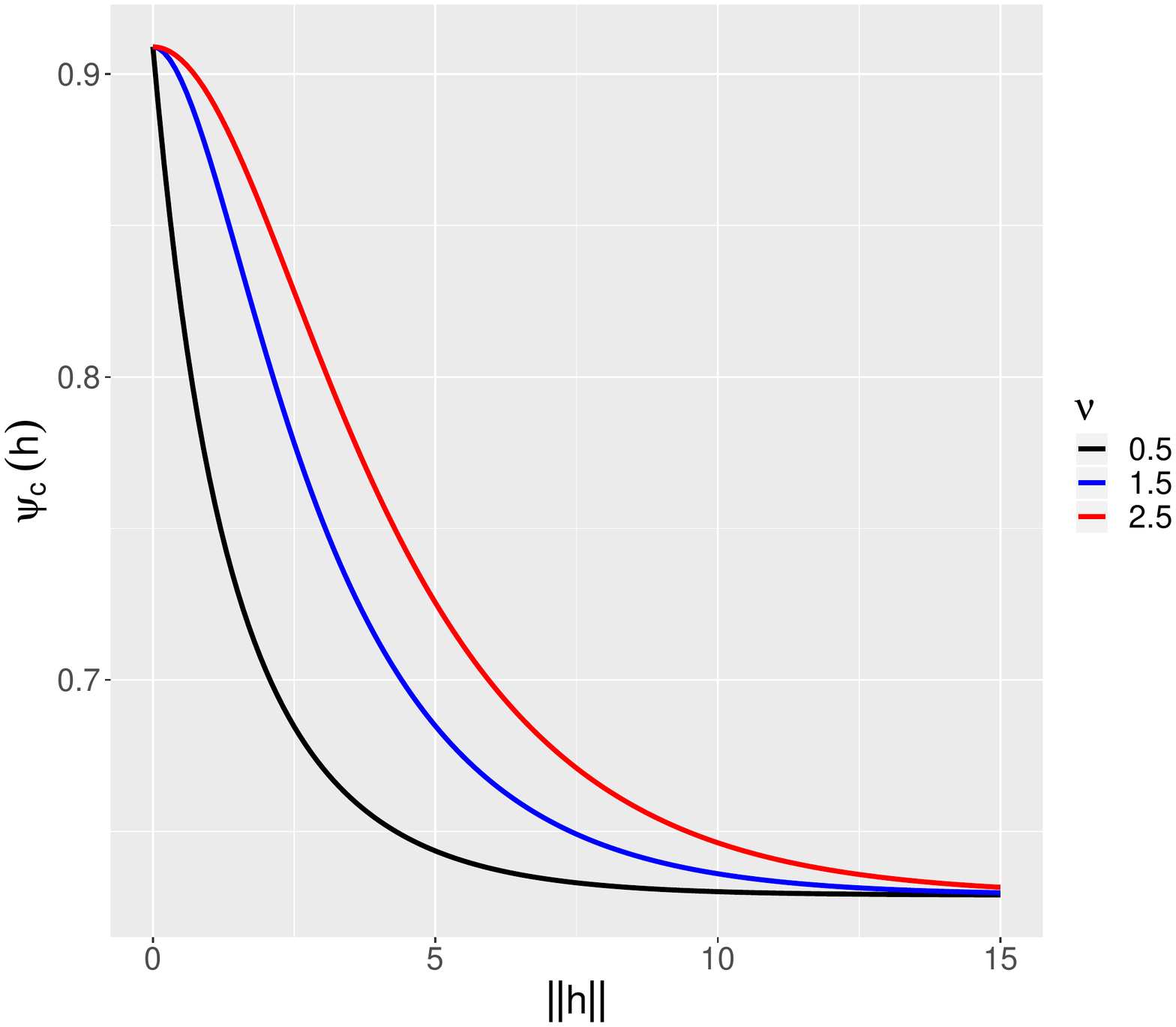}
\vspace{-10mm}
\caption{}
\end{subfigure}
\begin{subfigure}{1.0\textwidth}
  \centering
    \includegraphics[scale=0.3]{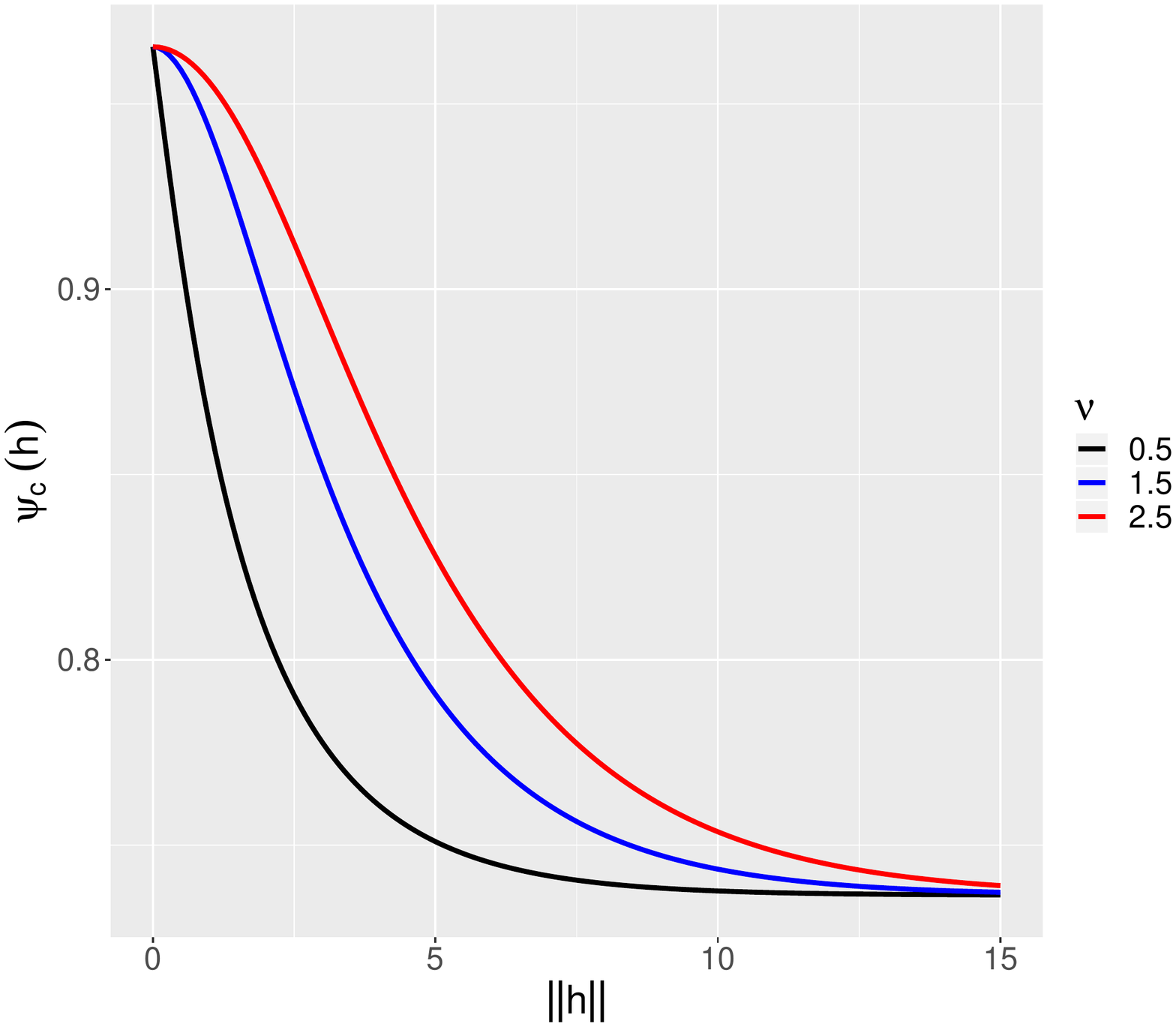}
\vspace{-10mm}
    \caption{}
\end{subfigure}
    \caption{$\psi_c(h)$ versus $h \in \{0,1, \ldots,15\}$. (a) $c=1.5$; (b) $c=2$; (c) $c=2.5$.}
    \label{fig:decaying1}
\end{figure}

\begin{figure}[hpt]
\centering
\includegraphics[scale=0.8]{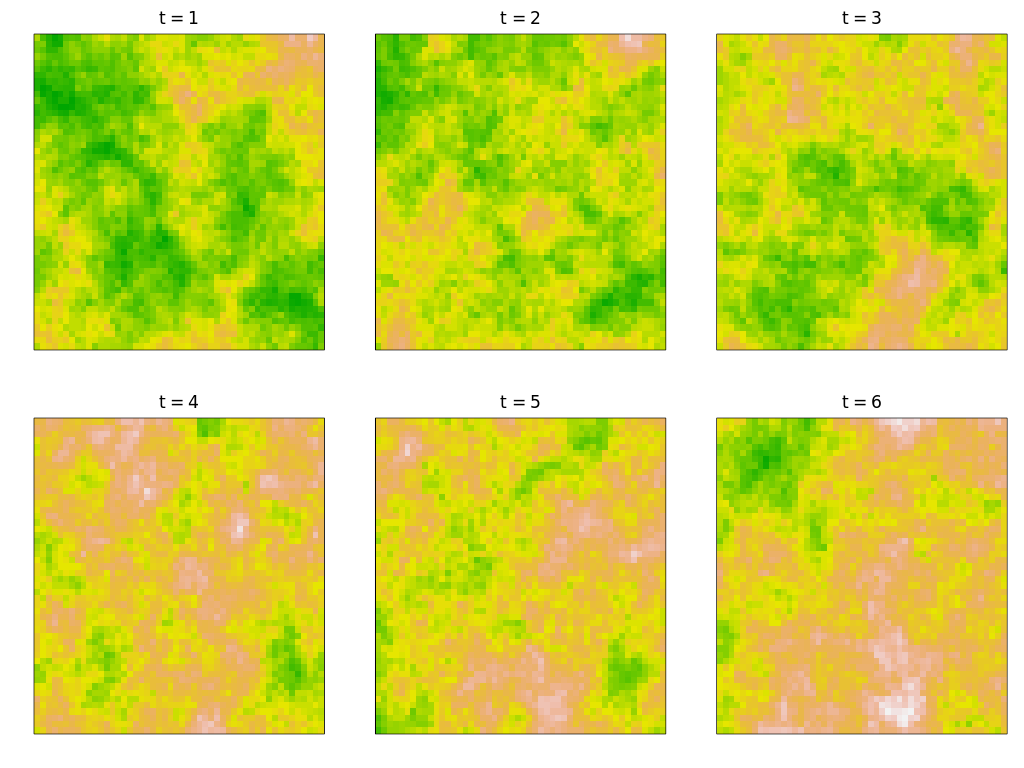}
\caption{Simulated realizations of a spatiotemporal process defined by a Gaussian random field with an exponential separable covariance function for $N_S=50$ and $N_T =6$.}
\label{fig:realisation1}
\end{figure}

\begin{figure}[hpt]
\centering
\includegraphics[scale=0.8]{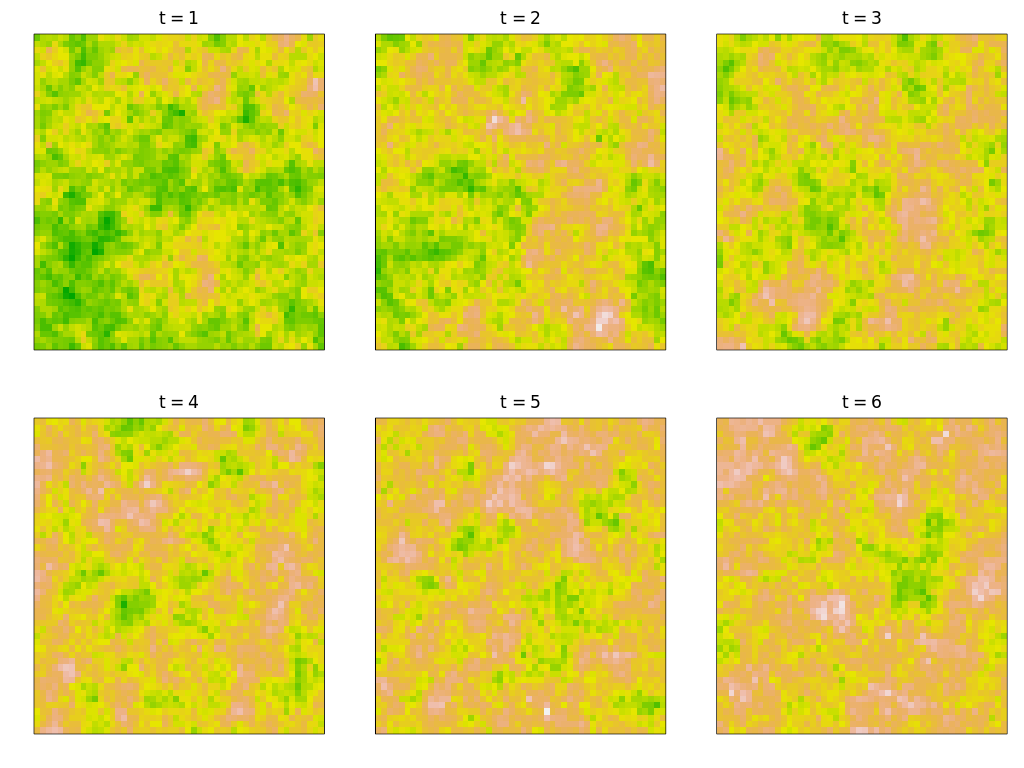}
\caption{Simulated realization of a spatiotemporal process defined by a Gaussian random field with an Iacosecare non-separable covariance function for $N_S=50$ and $N_T =6$.}
\label{fig:realisation2}
\end{figure}

\begin{figure}[hpt]
\centering
\includegraphics[scale=0.82]{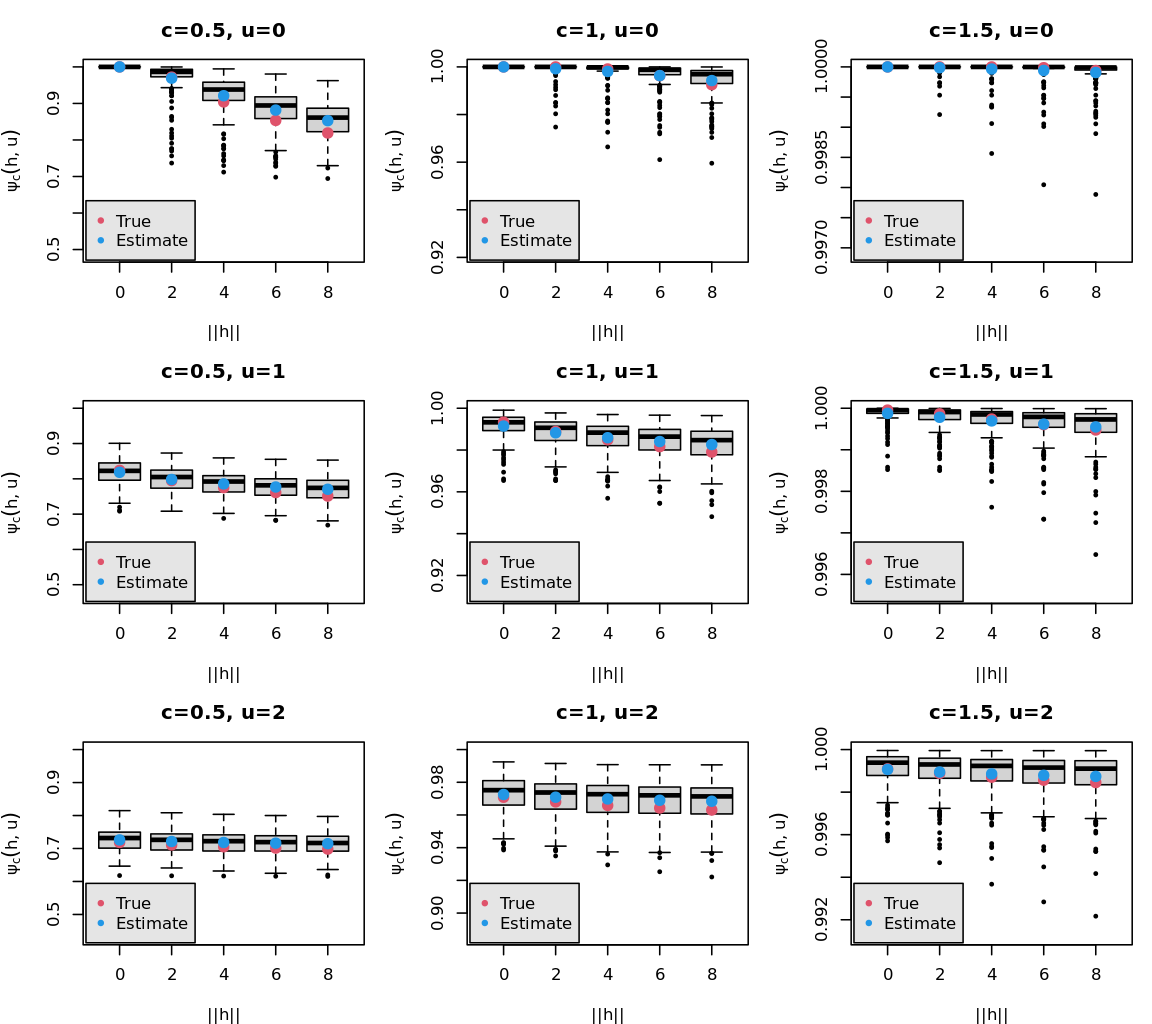}
\caption{Estimates of the probability of agreement as a function of $\|\bm h\|$, $u$ and $c$ for a spatiotemporal Gaussian process with a negative linear trend and an exponential separable covariance structure with fixed parameters given in Table 1 (in the manuscript), and for $N_S=20$. Note differences in range limits of the \textit{y}-axis among the nine panels.}
\label{fig:psi_st_separable1_ns20}
\end{figure}

\begin{figure}[hpt]
\centering
\includegraphics[scale=0.82]{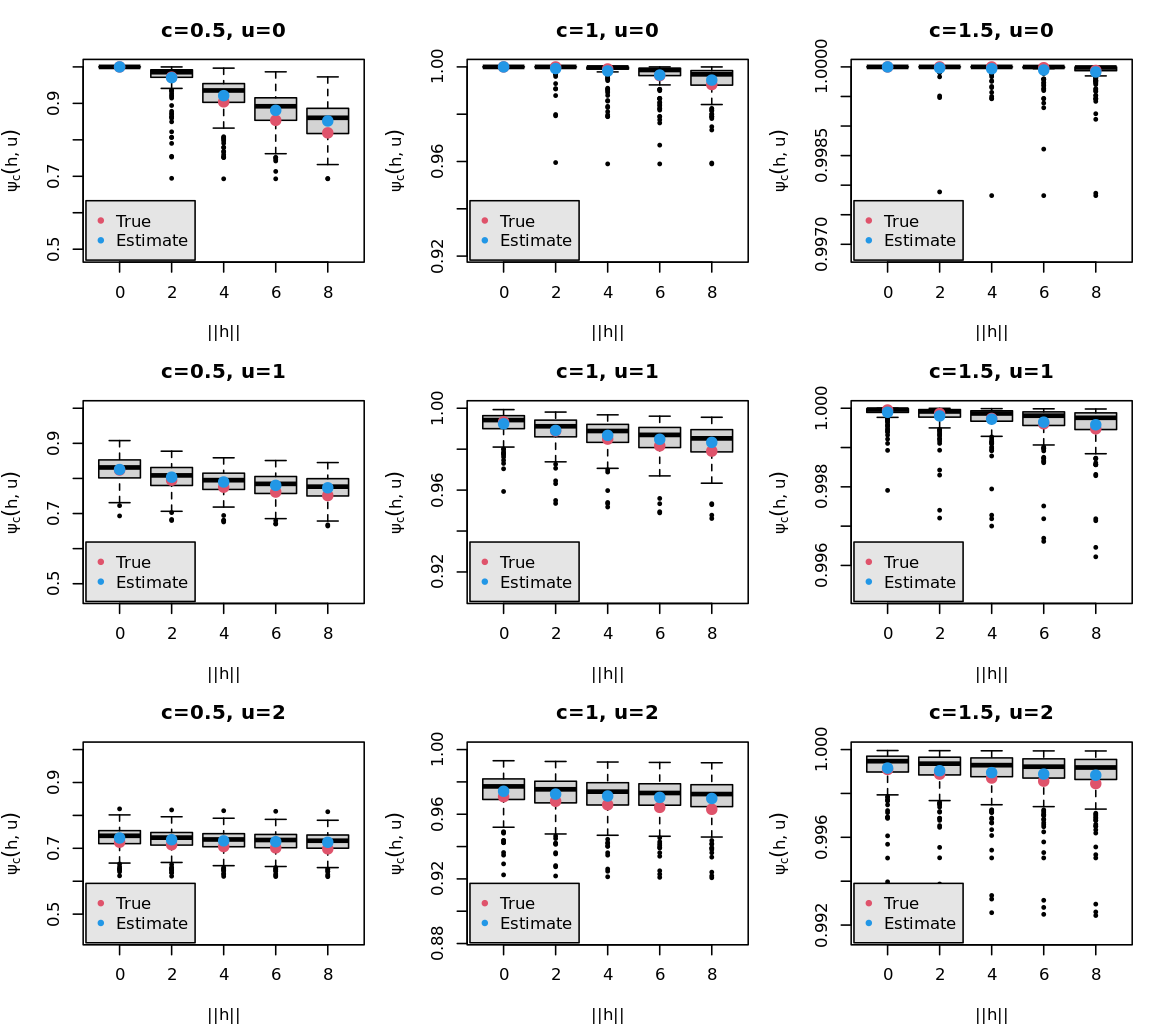}
\caption{Estimates of the probability of agreement as a function of $\| \bm h\|$, $u$ and $c$ for a spatiotemporal Gaussian process with a positive linear trend and an exponential separable covariance structure with fixed parameters given in Table 1 (in the manuscript), and for $N_S=20$. Note differences in range limits of the \textit{y}-axis among the nine panels.}
\label{fig:psi_st_separable2_ns20}
\end{figure}

\begin{figure}[hpt]
\centering
\includegraphics[scale=0.82]{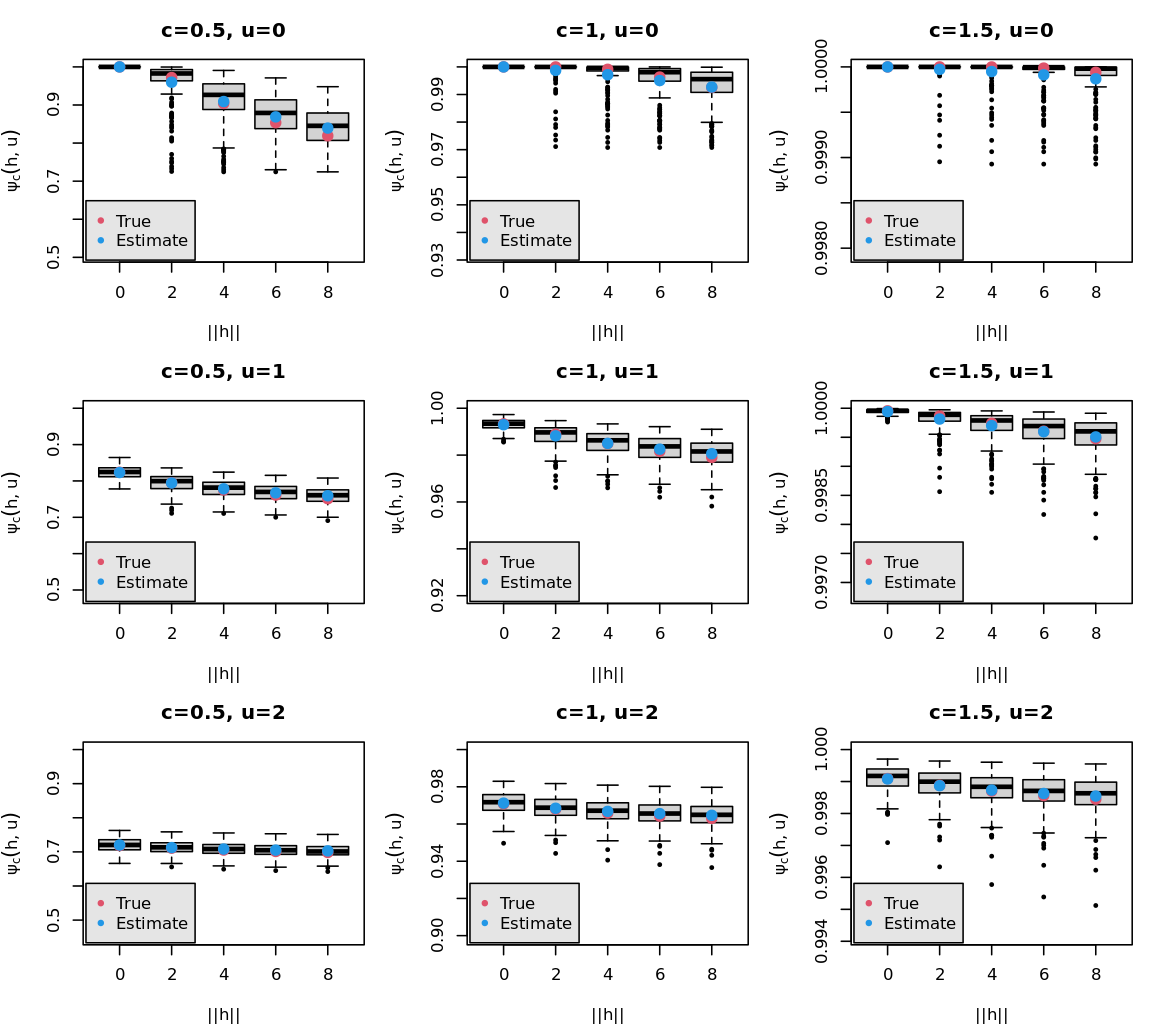}
\caption{Estimates of the probability of agreement as a function of $\|\bm h\|$, $u$ and $c$ for a spatiotemporal Gaussian process with a negative linear trend and an exponential separable covariance structure with fixed parameters given in Table 1 (in the manuscript), and for $N_S=50$. Note differences in range limits of the \textit{y}-axis among the nine panels.}
\label{fig:psi_st_separable1_ns50}
\end{figure}

\begin{figure}[hpt]
\centering
\includegraphics[scale=0.82]{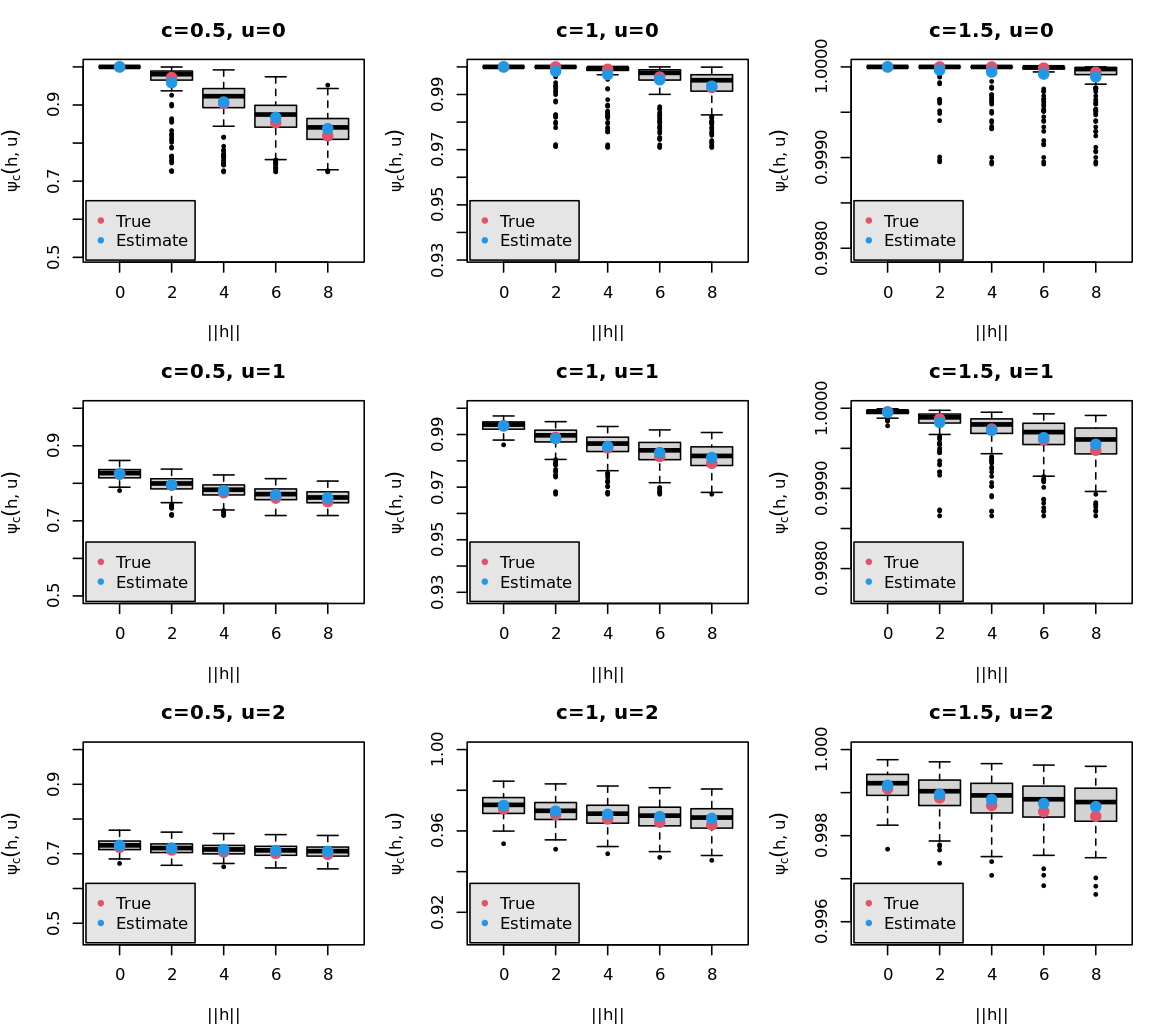}
\caption{Estimates of the probability of agreement as a function of $\| \bm h\|$, $u$ and $c$ for a spatiotemporal Gaussian process with a positive linear trend and an exponential separable covariance structure with fixed parameters given in Table 1 (in the manuscript), and for $N_S=50$. Note differences in range limits of the \textit{y}-axis among the nine panels.}
\label{fig:psi_st_separable2_ns50}
\end{figure}

\begin{figure}[hpt]
\centering
\includegraphics[scale=0.82]{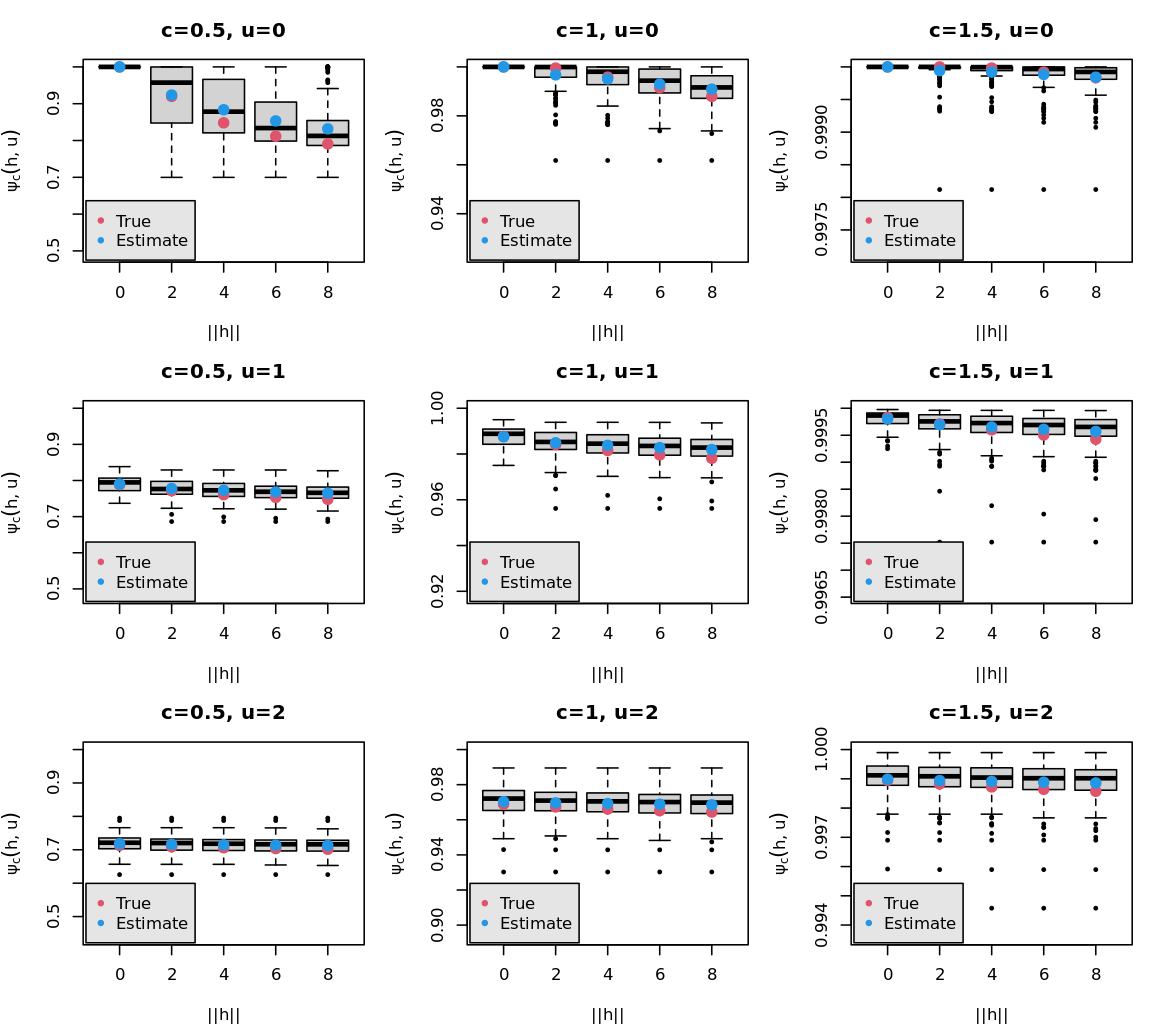}
\caption{Estimates of the probability of agreement as a function of $h$, $u$ and $c$ for a spatiotemporal Gaussian process with a negative linear trend and a non-separable Iacocesare covariance structure with fixed parameters given in Table 1 (in the manuscript), and for $N_S=20$. Note differences in range limits of the \textit{y}-axis among the nine panels.}
\label{fig:psi_st_nonseparable1_ns20}
\end{figure}

\begin{figure}[hpt]
\centering
\includegraphics[scale=0.82]{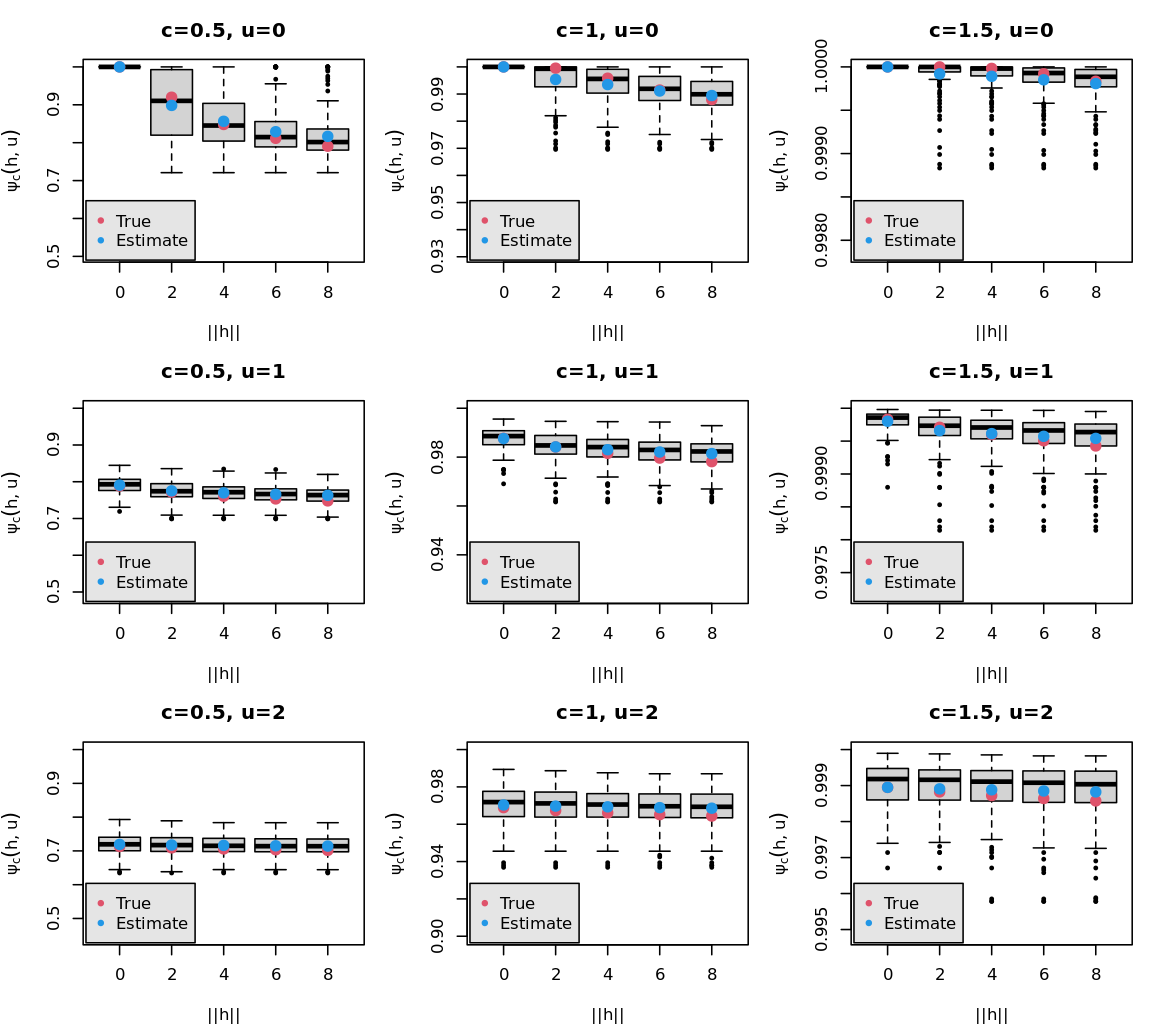}
\caption{Estimates of the probability of agreement as a function of $\|\bm h\|$, $u$ and $c$ for a spatiotemporal Gaussian process with a negative linear trend and a non-separable Iacocesare covariance structure with fixed parameters given in Table 1 (in the manuscript), and for $N_S=20$. Note differences in range limits of the \textit{y}-axis among the nine panels.}
\label{fig:psi_st_nonseparable2_ns20}
\end{figure}

\begin{figure}[hpt]
\centering
\includegraphics[scale=0.82]{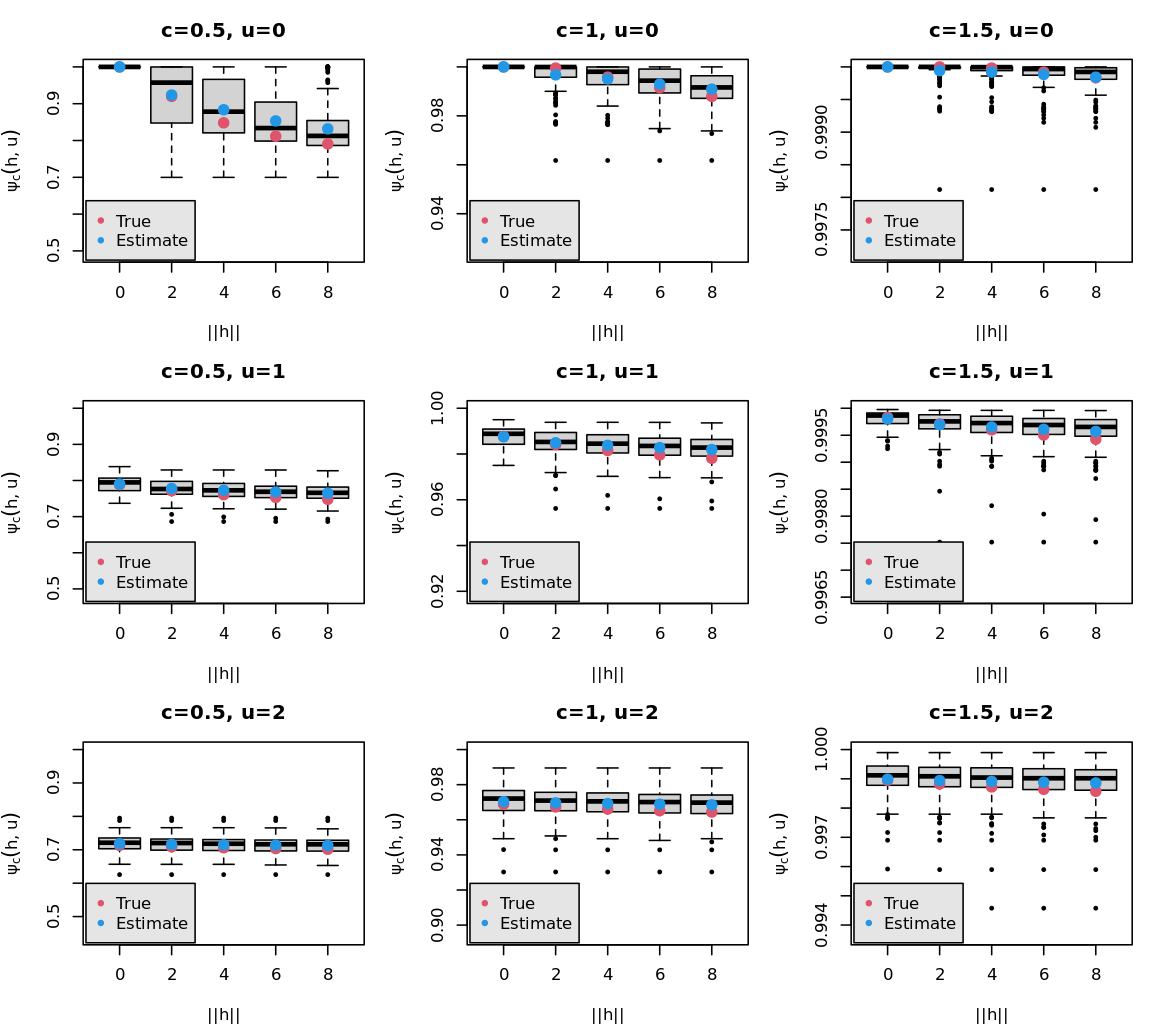}
\caption{Estimates of the probability of agreement as a function of $\|\bm h\|$, $u$ and $c$ for a spatiotemporal Gaussian process with a negative linear trend and a non-separable Iacocesare covariance structure with fixed parameters given in Table 1 (in the manuscript), and for $N_S=50$. Note differences in range limits of the \textit{y}-axis among the nine panels.}
\label{fig:psi_st_nonseparable1_ns50}
\end{figure}

\begin{figure}[htp]
	\centering
	\includegraphics[width=.9\textwidth]{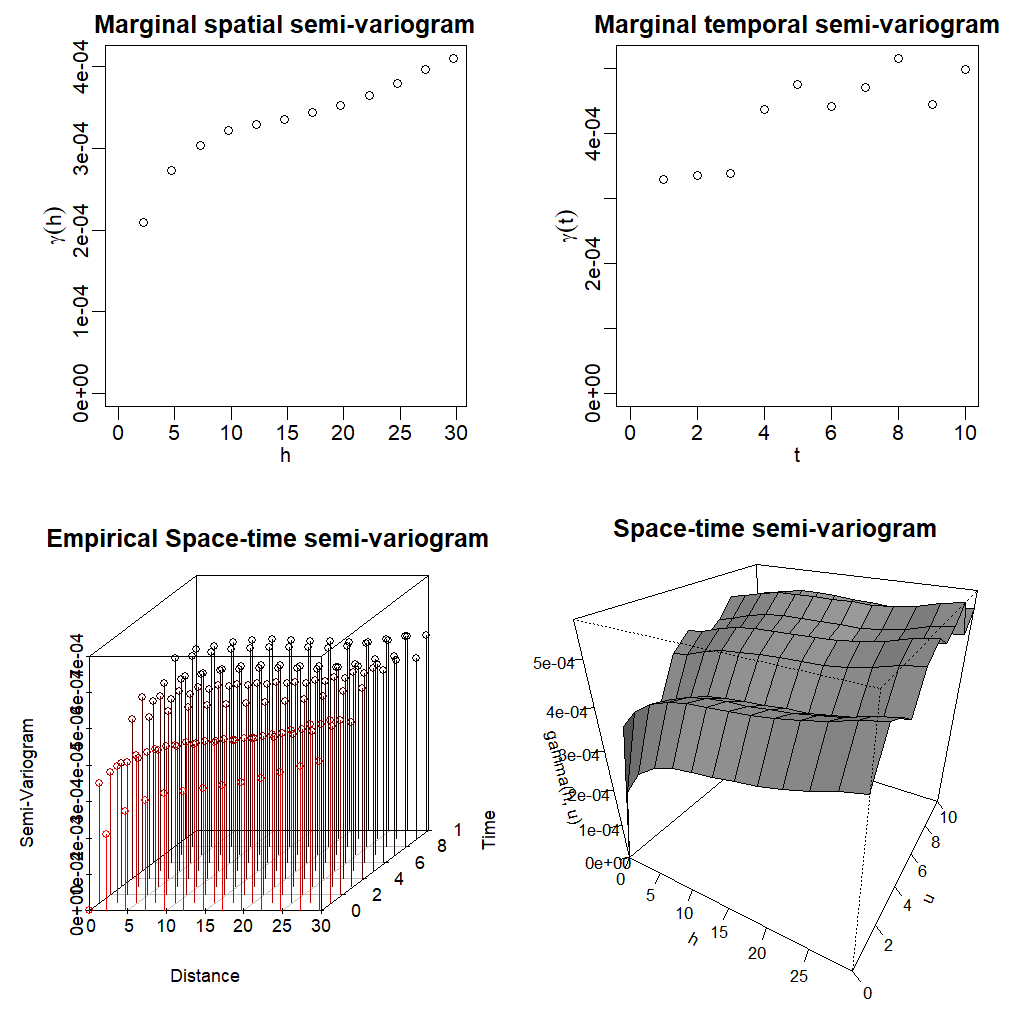}
	\caption{Empirical description of the data set.}
	\label{fig:variogrmas_st}
\end{figure}


\end{document}